\documentclass{aa}  
\usepackage{color} 

\usepackage{graphicx}
\usepackage{subfigure}
\usepackage{txfonts}
\usepackage[pdftex]{hyperref}
\begin{document}

   \title{Pulsational instabilities  driven by the $\epsilon$ mechanism in hot pre-horizontal branch stars}

   \subtitle{I.  The Hot-Flasher Scenario }

   \author{Tiara Battich\inst{1,2}, Marcelo M. Miller Bertolami\inst{1}, Alejandro H. C\'orsico\inst{1,2} \& Leandro G. Althaus\inst{1,2} }

   \institute{Instituto de Astrof\'isica de La Plata, CONICET--UNLP, Argentina
  		\and     
     		Facultad de Ciencias Astron\'omicas y Geof\'{\i}sicas, UNLP, Argentina \\
            \email{tbattich@fcaglp.unlp.edu.ar}}

   \date{Received M D, Y; accepted M D, Y}
\authorrunning{Battich et al.}
 
  \abstract
   {The $\epsilon$ mechanism is a self-excitation mechanism of stellar pulsations which acts in regions inside the star where nuclear burning takes place. It has been shown that the $\epsilon$ mechanism can excite pulsations in models of hot pre-horizontal branch stars before they settle into the stable helium core-burning phase. Moreover, it has been shown that the shortest periods of LS IV-14$^{\circ}$116, a mild He-sdBV star, could be explained that way.}
   {We aim to study the $\epsilon$ mechanism in stellar models appropriate for hot pre-horizontal branch stars to predict their pulsational properties and the instability domain in the $\log g-\log T_{\rm eff}$ plane.}
   {We perform detailed computations of non-adiabatic non-radial pulsations on stellar models during the helium subflashes just before the helium-core burning phase. Computations are carried out for different values of initial helium composition, metallicity and envelope mass at the moment of helium flash.}
   {We find an instability domain of long-period gravity modes due to the $\epsilon$ mechanism in the $\log g-\log T_{\rm eff}$ plane at roughly $22000\,{\rm K} \lesssim T_{\rm eff}\lesssim 50000\,$K and $4.67  \lesssim \log g \lesssim 6.15$. Consequently, we find instabilities due to the $\epsilon$ mechanism on pre-extreme horizontal branch stellar models ($T_{\rm eff}\gtrsim 22000\,$K), but not on
pre-blue horizontal branch stellar models ($T_{\rm eff}\lesssim 21000\,$K). The periods of excited modes range between $\sim 200$ and $\sim2000\,$s. Comparison with the three known pulsating He-rich subdwarfs shows that $\epsilon$ mechanism can excite gravity modes in stars with similar surface properties (He-abundances, $\log g$, $\log T_{\rm eff}$) but in our models it is only able to excite modes in the range of the shortest observed periods.}
   {We predict a new instability strip for hot-subdwarf stars of which  LS IV-14$^{\circ}$116 could be the first inhabitant.  Based on simple estimates we expect 1 to 10 stars in the current samples of hot-subdwarf stars to be pulsating by the $\epsilon$ mechanism. Our results could constitute a theoretical basis for future searches of pulsators in the Galactic field.}


   \maketitle
%

\section{Introduction}

Low-mass stars during the core helium (He)-burning phase of stellar evolution are located in a well defined region of the
Hertzsprung-Russell (HR) diagram called the Horizontal Branch (HB). The HB derives its name from its locus in the theoretical HR diagram, which spans a much thinner range in luminosity than in effective temperature. Some globular clusters present a population of HB stars hotter than the RR-Lyrae instability strip. This hot part of the HB is called the blue horizontal branch (BHB) and is located approximately in the effective temperature ($T_{\rm eff}$) range of $7000\,{\rm K}\lesssim T_{\rm eff}\lesssim21000\,{\rm K}$.  Some clusters also show an extension of the HB at even higher temperatures ($T_{\rm eff}\gtrsim22000K$) called the extreme horizontal branch (EHB). These EHB stars correspond to the hot-subdwarf stars of spectral type B (sdB), that can also be found in the Galactic Disk \citep{2016PASP..128h2001H}. The $T_{\rm eff}$ of stars in the BHB or the EHB depend mainly on the mass of their hydrogen (H)-rich envelopes. EHB stars (or sdBs) are core He-burning stars with H envelopes too thin to sustain H burning \citep{2016PASP..128h2001H}. In low-mass stars the beginning of the He-core burning phase occurs in the form of a violent He-core flash. This happens because in these stars He ignition occurs under degenerate conditions of the stellar plasma. After the main He-core flash, subsequent subflashes take place, although with less nuclear energy release than in main He flash (see \citealt{2012sse..book.....K}). Finally, He burning becomes stable and the stars settle into the Zero Age Horizontal Branch (ZAHB).  In this paper we call pre-EHB (pre-BHB) stars to stars undergoing the He subflashes before the ZAHB and with temperatures corresponding to the EHB (BHB).

The atmospheres of sdB stars are mostly H pure due to the action of gravitational settling, that carries the elements heavier than H to the interior of the star. However, there are some sdBs with He-enriched atmospheres (He-sdB). The short duration ($\sim 1.5-2\,$Myr) of the pre-EHB phase along with the action of stellar winds would prevent the formation of a H-pure envelope in this evolutionary stage. It has been argued that the He-sdBs could be stars during the He core subflashes just before the beginning of the quiescent He-core burning phase \citep{2010MNRAS.409..582N}. Hot-subdwarf stars of spectral type O (sdO) have a less clear evolutionary status. In contrast to the sdB stars, their atmospheres show a wide range of He abundances, and their temperatures are higher ($37000\,{\rm K}\lesssim T_{\rm eff} \lesssim 70000\,$K). The sdO population is believed to be a mixture of post-EHB stars, post-asymptotic giant branch stars, the progeny of double He-core white dwarf (He-WD) mergers, and  post red-giant branch (RGB) stars which had underwent a late He-core flash \citep{2009ARA&A..47..211H,2016PASP..128h2001H}. In particular, the He-rich sdO stars (He-sdO) are thought to be formed by double He-WD mergers \citep{2012MNRAS.419..452Z} and late helium-core flashes \citep{2008A&A...491..253M}. If the last scenario is holding, a fraction of the He-sdOs could also be stars undergoing He subflashes.
 
Among sdB stars, some of them exhibit periodic photometric variations likely due to global pulsations. There are two main classes of known pulsating stars among sdB stars, the slow pulsators (sdBVs or V1093 Her stars, \citealt{2003ApJ...583L..31G}) with long periods ($\sim 2500-8000\,$s) associated to gravity ($g$) modes, and the rapid pulsators (sdBVr or V361 Hya stars, \citealt{1996ApJ...471L.103C,1997MNRAS.285..640K}) with short periods ($\sim 80-400\,$s) associated with radial and non-radial pressure ($p$) modes \citep{2010IBVS.5927....1K}. In both cases the pulsational instabilities are explained by means of the $\kappa$ mechanism acting on the partial ionization zone of the iron-group \citep{1997ApJ...483L.123C,2003ApJ...597..518F}. Apart from the pulsating sdB stars, there exist also some sdO stars that exhibit variability \citep{2006MNRAS.371.1497W,2010MNRAS.401...23R,2016A&A...589A...1R}. Pulsating sdO stars, the most of which belong to the globular cluster $\omega$ Cen, are rapid pulsators with periods of $\sim 60-130\,$s, and are also understood by means of the $\kappa$ mechanism \citep{2008A&A...486L..39F,2010MNRAS.402..295R,2016A&A...589A...1R}. However, there are a few hot-subdwarf stars, LS IV-14$^\circ$116, KIC 1718290 and UVO 0825+15,  whose pulsations cannot be explained by means of the $\kappa$ mechanism. All these stars have in common the peculiarity of being mild He-enhanced hot-subdwarf stars. LS IV-14$^\circ$116 is a He-sdB star whose pulsations have periods in the range of sdBVs stars but its $\log g$ and $\log T_{\rm eff}$ values correspond to that of sdBVr stars \citep{2005A&A...437L..51A,2011ApJ...734...59G}. KIC 1718290 is a cool sdB star or a hot BHB star ($T_{\rm eff} \sim$ 22100 K, \citealt{2012ApJ...753L..17O}). Its effective temperature is close to the sdBVs stars and its pulsation modes are compatible with long-period $g$ modes, but has periods that are too long compared with those of the sdBVs stars (up to 11 h, \citealt{2012ApJ...753L..17O}). Finally, \citet{2017MNRAS.465.3101J} recently reported variability in the light curve of UVO 0825+15, a He-sdO star, compatible with non-radial pulsations associated to long-period $g$ modes. UVO 0825+15 is located outside the instability region of sdBVs stars and its periods are also of the order of a few hours ($\sim 12$ h).

\citet{2011ApJ...741L...3M} suggested that the pulsations of LS IV-14$^\circ$116 could be due to $\epsilon$ mechanism acting during the He-subflashes before the star settles into the quiescent He-burning phase. The $\epsilon$ mechanism is a self-excitation mechanism of stellar pulsations which acts on the regions inside the star where nuclear burning takes place. Due to the strong dependence of nuclear burning with the temperature, even tiny perturbations of the temperature inflicted by oscillations translate into huge increases in the amount of nuclear energy released, which in turn increases the local temperature. The process is fed back, resulting in a global instability that grows with time. Generally, the $\epsilon$ mechanism is not a very efficient mechanism of excitation since the pulsation amplitudes tend to be small in the high-temperature layers where nuclear energy is generated. The $\epsilon$ mechanism was studied in a variety of scenarios, including white dwarf (WD) stars \citep{1997ApJ...489L.149C, 2014ApJ...793L..17C,2016A&A...585A...1C,2016A&A...595A..45C}, pre-WD stars \citep{1986ApJ...306L..41K,1996ASPC...96..361S,1997A&A...320..811G,2009ApJ...701.1008C,2014PASJ...66...76M}, post He-WD mergers \citep{2013EPJWC..4304004M}, post-main sequence B stars \citep{2012ApJ...749...74M}, and main sequence low-mass stars \citep{2005A&A...432L..57P,2012MNRAS.419L..44R,2012MNRAS.422.2642S} among others. All these studies predict instabilities due to the $\epsilon$ mechanism. Indeed, this mechanism was proposed as the responsible of the pulsations of a few stars, as the supergiant B star Rigel \citep{2012ApJ...749...74M}, the PNNV star VV 47 \citep{2006A&A...454..527G,2009ApJ...701.1008C} and, as mentioned, de He-sdB star LS IV-14$^\circ$116 \citep{2011ApJ...741L...3M}, although none of these suggestions are conclusive. Therefore, if the pulsations of LS IV-14$^\circ$116, or another He-enhanced hot-subdwarf star, are confirmed to be triggered by $\epsilon$ mechanism, it would represent the first evidence that $\epsilon$ mechanism can indeed excite pulsations in stars. 

Moreover, observational proof of the existence of the He flash and subflashes is lacking, as pointed out by \citet{2012ApJ...744L...6B}. Particularly, there is no hint of any star passing through the He-flash or subflashes. \citet{2009A&A...501..659M} performed 3D and long term (36 h) 2D simulations of the evolution of the core He flash. They found that the convective zone that develops as a result of the He burning grows rapidly, and argue that if it continues to grow at the same rate for $\sim$ a month, the convection would eventually lift the electron degeneracy and the He-flash would not be followed by subsequent subflashes. In this context, an observational hint of the existence of the subflashes would be very interesting. If any star is identified as pulsating by $\epsilon$ mechanism in a pre-EHB phase, it would represent a first proof of the existence of He-subflashes.
 
Therefore, faced with the possibility of $\epsilon$ mechanism being responsible for the pulsations in LS IV-14$^\circ$116, along with the recent discovery of new long-period pulsating He-enriched hot-subdwarf stars, it is of interest to have a detailed study of the pulsations triggered by the $\epsilon$ mechanism on pre-BHB and pre-EHB model stars. Such a study is lacking, since \citet{2011ApJ...741L...3M} studied just a single evolutionary sequence. In this work we largely extend the study of \citet{2011ApJ...741L...3M} by performing a comprehensive stability analysis on stellar models appropriate for stars on the pre-BHB and pre-EHB, that undergo the He subflashes. We extend the previous work of \citet{2011ApJ...741L...3M} by computing models of different hot-flashers
flavours and different initial compositions, and performing
non-adiabatic pulsation studies for the whole range of temperatures of the hot-horizontal branch. This allows us to determine the domain of
instability due to the $\epsilon$ mechanism. In addition we provide a detailed
theoretical study of the excitation and damping of the oscillations and provide the expected growth rates of the modes as well as the expected rates of period changes. This will allow to make better comparisons with the available and future observations. Moreover, our results could constitute a theoretical basis for guide future searches of pulsating stars of this kind.

The paper is organized as follows. In Sect. \ref{sec:num} we present the input physics and numerical tools of the simulations.  
In Sect. \ref{sec:evol} we discuss the evolutionary sequences in the hot-flasher scenario, their predicted surface abundances, and compare them with previous works. In Sect. \ref{sec:results} we present in detail the results of the stability analysis. In Sect. \ref{sec:disc} we compare the results with the available observations and give a brief discussion. Finally, in Sect. \ref{sec:conclusion} we summarize our conclusions.

\section{Input physics and numerical tools}
\label{sec:num}

All the stellar evolutionary calculations in this work were performed with the \texttt{LPCODE} stellar evolutionary code. The \texttt{LPCODE} is a well tested code \citep{2013A&A...555A..96S,2016A&A...588A..25M} that has been used for a variety of studies regarding low-mass stars and WDs (see e.g. \citealt{2010Natur.465..194G,2011A&A...533A.139W,2016ApJ...823..158C} ). 
In particular, \texttt{LPCODE} was used in the computation of He-flashes, including the born again episode \citep{2006A&A...449..313M} and the hot-flasher scenario \citep{2008A&A...491..253M}. Recently it has been used to develop a new grid of post-asymptotic giant branch models \citep{2016A&A...588A..25M} and study the evolution of WDs originated from He-enhanced, low-metallicity progenitors \citep{2017A&A...597A..67A}. The \texttt{LPCODE} is well described in \citet{2003A&A...404..593A,2005A&A...435..631A}, and the improvements of the last version are detailed in \citet{2016A&A...588A..25M}. Next, we mention the relevant input physics for this work. 

The nuclear network of \texttt{LPCODE} accounts for 16 elements along with 34 nuclear reactions for the p-p chains, CNO bi-cycle, He burning and carbon ignition. The reactions rates are the same as in \citet{2005A&A...435..631A} with the exception of the reactions $^{12}{\rm C}+{p}\rightarrow ^{13}{\rm N} + \gamma \rightarrow ^{13}{\rm C}+e^++\nu_e$, $^{13}{\rm C}(p,\gamma) ^{14}{\rm N}$ and $^{14}{\rm N}(p,\gamma)^{15}{\rm O}$. The first two are taken from \citet{1999NuPhA.656....3A}, and the last one is taken from \citet{2005EPJA...25..455I}. Radiative opacities are those of OPAL \citep{1996ApJ...464..943I}, complemented at low-temperature with the molecular opacities of \citet{2005ApJ...623..585F} and \citet{2009A&A...508.1343W}. The conductive opacities are included according to \citet{2007ApJ...661.1094C}. The neutrino emission due to plasmon-emission process is calculated according to \citet{1994ApJ...425..222H}. Convection is solved within the standard mixing length theory (MLT). The free parameter of the MLT was chosen to be $\alpha_{\rm MLT}=1.822$ that correspond to solar calibration for the \texttt{LPCODE} \citep{2016A&A...588A..25M}. The mixing and burning processes are computed simultaneously in the context of diffusive convective mixing \citep{2005A&A...435..631A}. No extra-mixing processes were included in the simulations. 

All the stellar pulsation calculations were performed with the linear, non-radial, non-adiabatic stellar pulsation code \texttt{LP-PUL}, which is coupled to the \texttt{LPCODE}. The \texttt{LP-PUL} was widely used in studies of pulsation properties of low-mass stars (see e.g. \citealt{2006A&A...458..259C,2012MNRAS.420.1462R,2016A&A...588A..74C,2017A&A...597A..29S}). In particular it was used to study the $\epsilon$-mechanism effects on WD and pre-WD stars (\citealt{2009ApJ...701.1008C,2014ApJ...793L..17C,2016A&A...595A..45C}). The \texttt{LP-PUL} is fully described in \citet{2006A&A...458..259C} and references therein, with the inclusion of the $\epsilon$-mechanism mode driving as described in \citet{2009ApJ...701.1008C}. The non-adiabatic computations in this work rely on the "frozen-in convection" approximation, in which the perturbation of the convective flux is neglected. In addition, it was assumed that ${\rm d}S/{\rm d}t = 0$ in the non-perturbed background model adopted in the non-adiabatic computations. For a discussion of the validity of these assumptions in the study of the $\epsilon$ mechanism on the subflashes stage we refer the reader to the work of \citet{2011ApJ...741L...3M}.

\section{Evolutionary sequences}
\label{sec:evol}

In the canonical stellar evolution picture, the He core flash takes place when stellar models reach the tip of the RGB. However, under certain conditions, most of the H-rich envelope mass can be removed before the development of the He-core flash. These conditions could be enhanced winds due to stellar rotation \citep{1997fbs..conf....3S,2015Natur.523..318T}, mass transfer due to stable Roche lobe overflow or common envelope systems \citep{2003MNRAS.341..669H,1976IAUS...73...75P}, or the ingestion of a substellar companion \citep{2014A&A...570A..70S}. In addition, in He-enriched populations even standard winds can lead to the (almost) complete removal of the H-rich envelope before the RGB-tip in very low mass stars \citep{2012ApJ...748...62V, 2017A&A...597A..67A}. When this happens models depart from the RGB contracting towards higher effective temperatures at constant luminosity. \citet{1993ApJ...407..649C} demonstrated that a He flash can still develop when the model is entering the WD cooling stage. This scenario leads to He-core burning models with a wide range of envelope masses and thus  populating the hot end of the horizontal branch (\citealt{1972ApJ...173..401F}, and references therein).  \citet{1996ApJ...466..359D} introduced the term "hot-flashers" for this scenario.

We constructed pre-EHB and pre-BHB models within the hot-flashers scenario, i.e., removing different amounts of the H-rich envelope by an artificially enhanced mass loss at the RGB tip. We calculated the evolution of initially $1\,M_{\odot}$ models from the ZAMS to the RGB. At the tip of the RGB we switched on artificially-enhanced mass loss, removing different amounts of envelope mass. For the purpose of this work, the particular value of the mass at the ZAMS and the treatment of mass loss at RGB are not relevant, since it is the total mass at He ignition that determines the behaviour of the hot-flashers. Moreover, the actual mass loss rate can not be determined without a clear understanding of the exact process responsible for the formation of hot horizontal branch stars. 

Models were computed for three different initial compositions. The  adopted initial compositions are shown in Table \ref{table:2}. Two sets were computed with canonical initial He abundances according to the relation ${\rm Y}=2\,{\rm Z}+0.245$. Due to the evidences of He-enhanced populations in globular clusters (e.g. \citealt{2017arXiv170602278M}) we decided to calculate evolutionary sequences with ${\rm Y}=0.4$ to characterize the impact of He-enriched feature on $\epsilon$-mechanism driven pulsations on the hot subflashes stage. Next we describe the qualitative behaviour of hot-flashers and compare our sequences with other works.

\begin{table}
\caption{Initial abundances by mass fraction of the stellar sequences computed in this work.}              
\label{table:2}      
\centering                           
\begin{tabular}{c c c}          
\hline\hline     
X & Y & Z\\
\hline
0.752	& 0.247 & 0.001\\
0.695	& 0.285 & 0.02\\
0.58	& 0.4	&  0.02\\
\hline                             
\end{tabular}
\end{table}

\subsection{Qualitative behaviour of hot-flashers}
 
\begin{figure}
	\centering
	\includegraphics[width=0.48\textwidth]{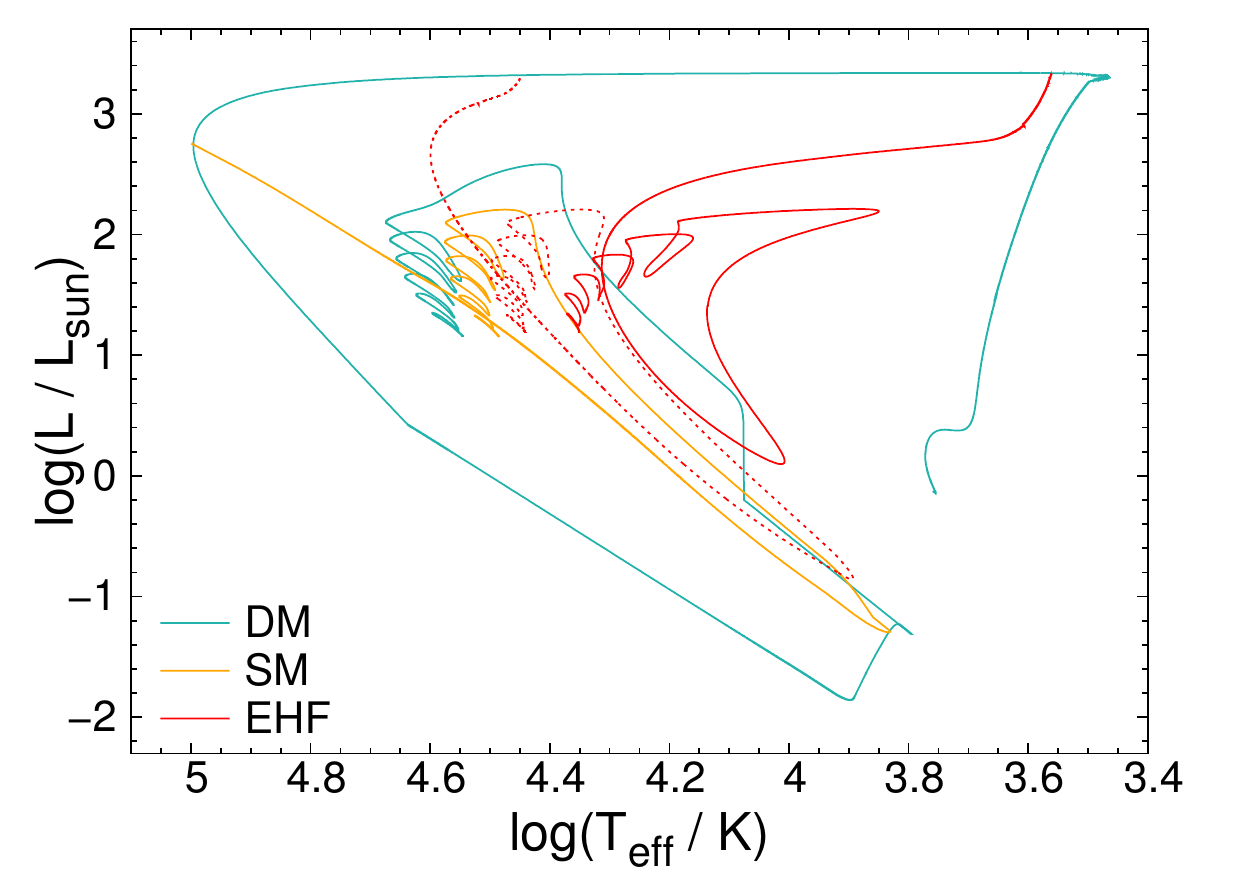}
	\caption{H-R diagram for evolutionary sequences with ${\rm Z}=0.02$. The evolutionary sequences correspond to two cases of early hot-flashers (EHF, dotted and full red lines) and two cases of late hot-flashers: one with shallow mixing (SM, full yellow line) and the other with deep mixing (DM, full light blue line).}
	\label{fig:hr:DM}
\end{figure}

\citet{2004ApJ...602..342L} classified the hot flashers into three cases, the early hot-flasher, the late hot-flasher with shallow mixing and the late hot-flasher with deep mixing (Fig. \ref{fig:hr:DM}). In the early hot-flasher (EHF), the He-flash occurs when the model star is moving from the tip of the RGB towards the top of the WD cooling curve. In this case, as in canonical He-core flashes, the convective zone that develops due to the He-core flash (He-core flash driven convective zone; HeFCZ) does not reach the H-rich envelope. This is a consequence of the existence of an entropy barrier at the H-burning shell \citep{1976ApJ...208..165I}. Consequently, the surface abundances are not altered during the He-flash 
but the models end up settling down at hotter effective temperatures than canonical HB models, due to their thinner H-rich envelopes. In late hot-flashers, the He flash takes place when the model stars are entering the WD cooling curve. As the model stars descend the WD cooling curve, the energy liberation by H shell burning decreases. Therefore also the entropy barrier decreases. At some point the H-burning shell becomes too weak to prevent the contact between the H-rich envelope and the He-flash driven convective zone. The penetration, or not, of the He-flash driven convective zone into the H-rich envelope divides the late hot flashers into deep-mixing and shallow-mixing events. Regardless the kind of late hot-flasher, after the He-flash, the stellar envelope expands due to the energy liberated by it, and the surface became cooler. As a consequence, an outer convective zone develops in the envelope which moves progressively inwards. In the shallow mixing (SM) case the He-core flash driven convective zone eventually splits giving rise to a convective zone that persists in the outer region of the core by several thousands years. This convective zone eventually merges with the outer convective zone of the envelope, and some material of the core is taken to the surface. In this case there are not H-burning, but the H surface abundance decreases.  
\citet{2008A&A...491..253M} found that in some cases the convective region that develops after He-flash can penetrate slightly into the H-rich material, and some H is burned. This case is labelled as SM*. In late-hot flasher with deep mixing (DM), the flash occurs where the model star is well down the WD cooling curve  and the entropy barrier is small. The convective region reaches the H-rich layer shortly after the maximum of He-flash, and the H is mixed within the hot HeFCZ \citep{1997fbs..conf....3S} leading to a violent burning of H as first calculated by \citet{2003ApJ...582L..43C}. This H-burning leads to a further outward growth of the convective zone, and more H is burned. Therefore, the burning is unstable and a H-flash develops, where almost all H is burned. The superficial abundances change drastically, and the envelope becomes highly He-enriched. The absence of H leads to even higher effective temperatures on the resulting ZAHB models.

With the EHF scenario we are able to populate the pre-BHB and the pre-EHB up to $T_{\rm eff}\simeq 37000\,$K. Late hot-flashers are able to produce models of even higher temperatures, up to $T_{\rm eff}\simeq 50000\,$K (depending on metallicity).

\subsection{Description of the sequences}

\begin{figure}
	\centering
	\includegraphics[width=0.48\textwidth]{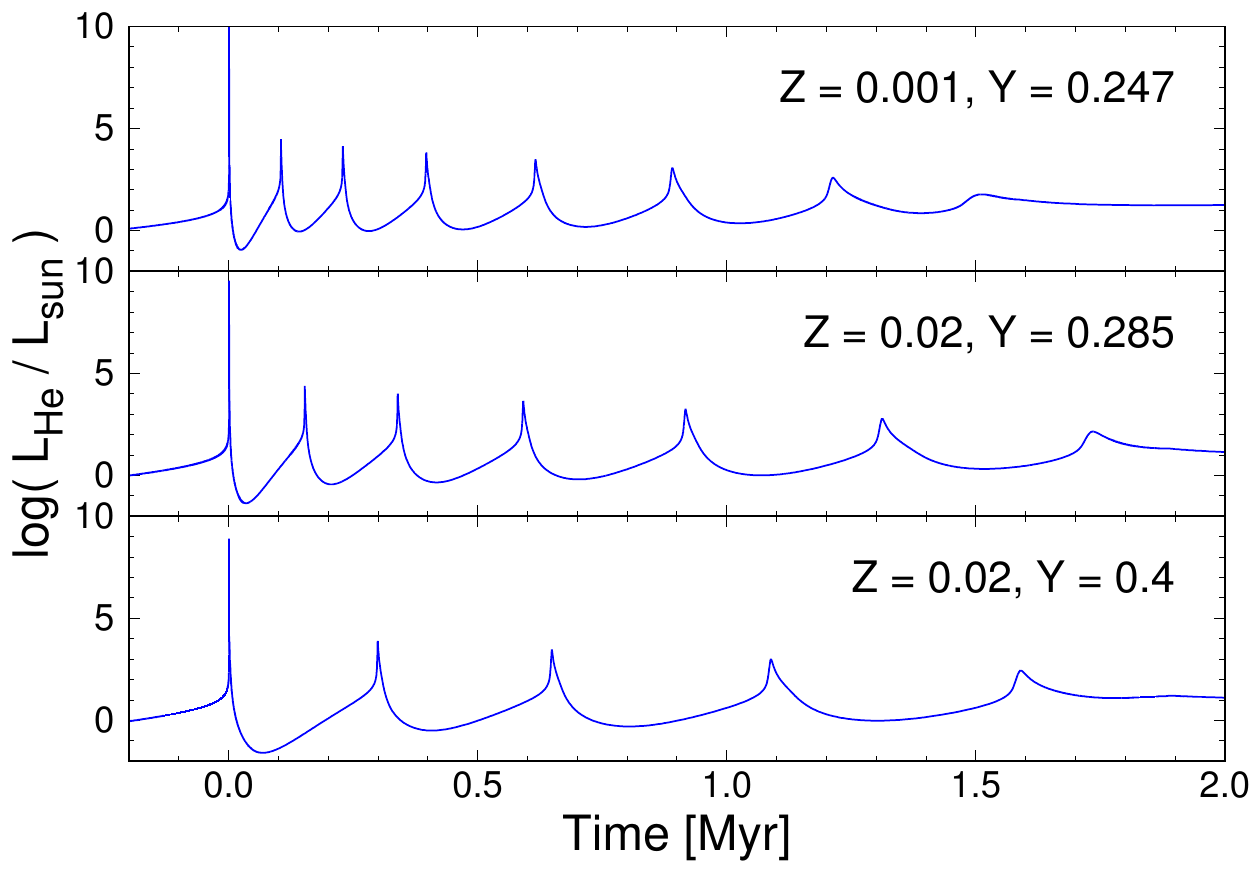}
	\caption{Luminosity due to nuclear energy liberation at the He-flash and subflashes vs. the time after the maximum energy liberation.}
	\label{fig:lum:time}
\end{figure}

In Fig. \ref{fig:lum:time} we show the typical evolution of the He-burning luminosity ($L_{\rm He}$) on He flash and subflashes for each choice of initial chemical compositions. As in canonical He-flashes \citep{2012sse..book.....K}, $L_{\rm He}$ reaches values as high as $L_{\rm He}\sim 1.3\times 10^{10}\,L_{\odot}$ ($L_{\rm He}\sim 3.6\times 10^{9}\,L_{\odot}$) for ${\rm Z}=0.001$ (${\rm Z}=0.02$) and standard He abundances. The energy released due to the He flash in the case of ${\rm Y}=0.4$ is a factor $\sim 4$ lower than in the cases with canonical initial He composition ($L_{\rm He}\sim 8.1\times 10^{8}\,L_{\odot}$). Also, the number of subflashes is lower for higher values of Z and Y. These trends are related to lower degree of degeneracy and lower core masses for higher values of Z and Y. In all cases, the time scale of the subflashes stage is about $\sim 2\,$Myr.

\begin{figure}
	\centering
	\subfigure{\includegraphics[width=0.45\textwidth]{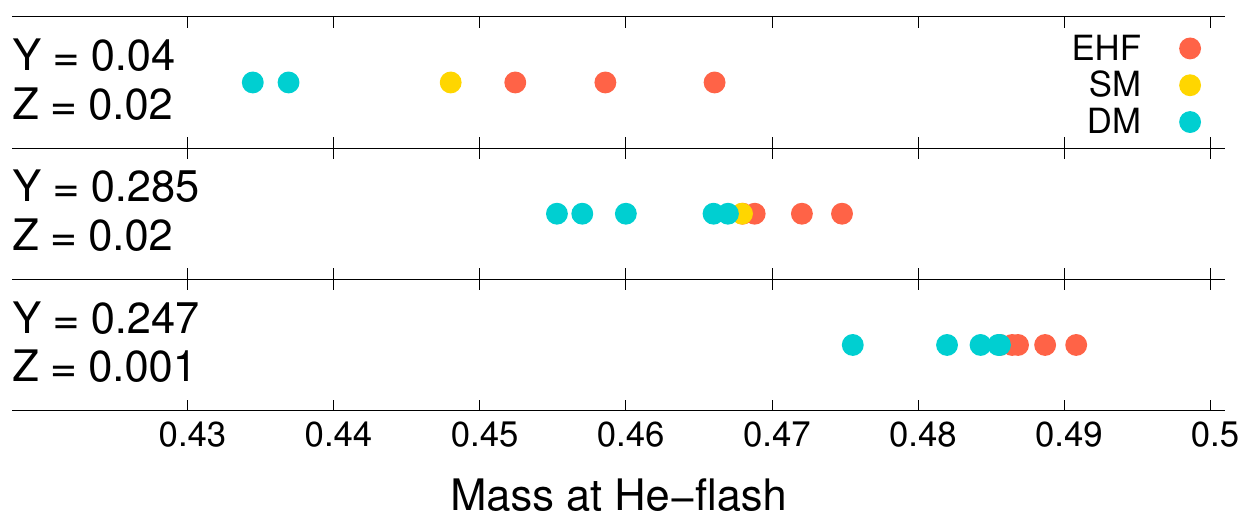}}
	\subfigure{\includegraphics[width=0.45\textwidth]{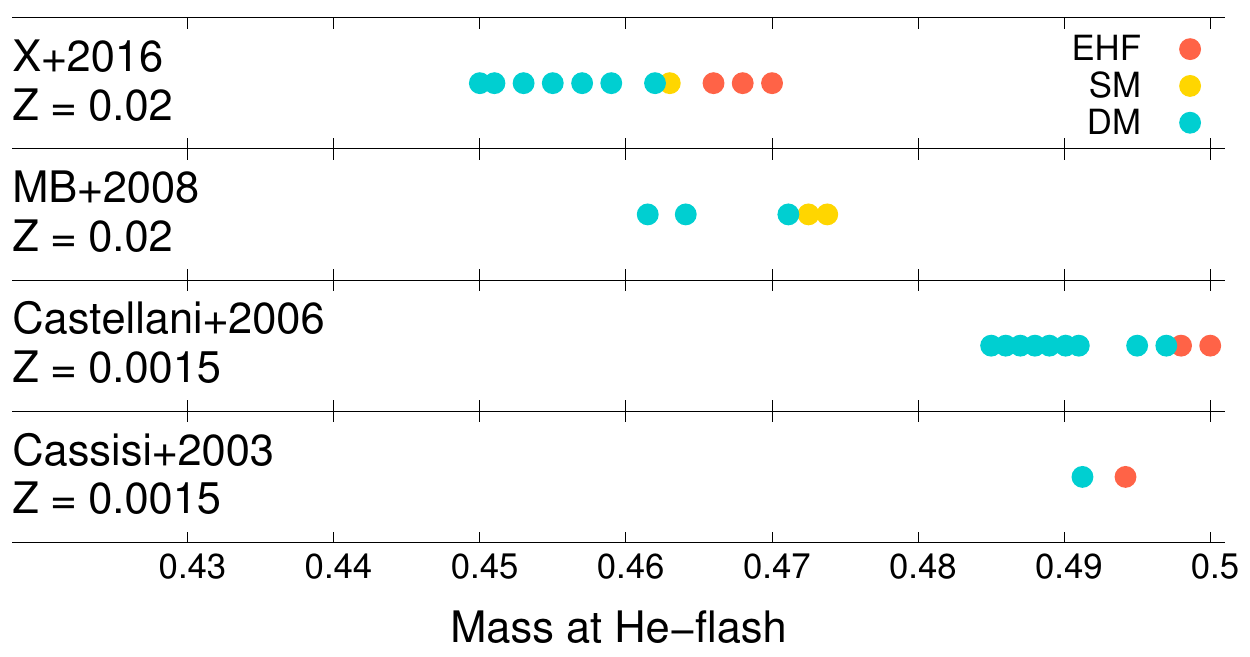}}
	\caption{Upper panel: mass of the stellar models at He-ignition for the different initial chemical compositions computed in this work. Lower panel: same as upper panel but for the works of \citet{2016arXiv160808739X} (X+2016) and \citet{2008A&A...491..253M} (MB+2008) for ${\rm Z}=0.02$; and \citet{2006A&A...457..569C} (Castellani+2006) and \citet{2003ApJ...582L..43C} (Cassisi+2003) for ${\rm Z}=0.0015$.}
\label{fig:metal:masa}
\end{figure}

Figure \ref{fig:metal:masa} shows the mass of the stellar models at the moment of He-ignition for each choice of initial chemical abundances. The He-core mass at He-ignition is lower for higher metallicities and higher He abundances.   
 Stars with He-enhanced abundances during the main sequence have a larger mean molecular weight. These stars are hotter and brighter for a given mass than stars with canonical He-contents. In particular, the higher core temperatures during the RGB phase mean that the temperature needed for He ignition is reached at lower He-core mass than in the case of normal He abundances \citep{2005essp.book.....S}. Due to the well known core-luminosity relation \citep{2012sse..book.....K}, hot-flasher models corresponding to sequences with higher initial He abundances are less luminous, and lie at higher gravities on the Kiel-diagram than their canonical counterparts.

The qualitative behaviour of our sequences is the same as in previous works \citep{1996ApJ...466..359D,2001ApJ...562..368B,2003ApJ...582L..43C,2006A&A...457..569C,2008A&A...491..253M}. In the lower panel of Fig. \ref{fig:metal:masa} we show the masses at He-flash obtained by \citet{2003ApJ...582L..43C} and \citet{2006A&A...457..569C} for ${\rm Z}=0.0015$, and by \citet{2008A&A...491..253M} and \citet{2016arXiv160808739X} for ${\rm Z}=0.02$. The masses of our stellar models at He-ignition for ${\rm Z}=0.001$ are lower than those of \citet{2003ApJ...582L..43C} and  \citet{2006A&A...457..569C}, in spite of their values corresponding to a higher metallicity. This is due to the adoption in our work of the updated conductive opacities of \citet{2007ApJ...661.1094C}. The masses of the hot-flasher sequences computed in this work for ${\rm Z}=0.02$ are similar of those obtained by \citet{2008A&A...491..253M} and \citet{2016arXiv160808739X} for the same metallicity, being the \citet{2016arXiv160808739X} values slightly lower and the \citet{2008A&A...491..253M} ones slightly higher than in the present work. \citet{2016arXiv160808739X} compute the formation of sdB stars within the ``common-envelope ejection channel''. They mimic the envelope ejection just by enhancing the mass loss at the RGB-tip. Since the behaviour of the hot-flasher sequences depends only on the remaining  envelope mass, and not on the detailed treatment of mass loss, their evolutionary sequences also belong to the hot-flasher scenario, and are comparable with other hot-flasher scenario sequences. 

\begin{figure}
	\centering
	\subfigure{\includegraphics[width=0.45\textwidth]{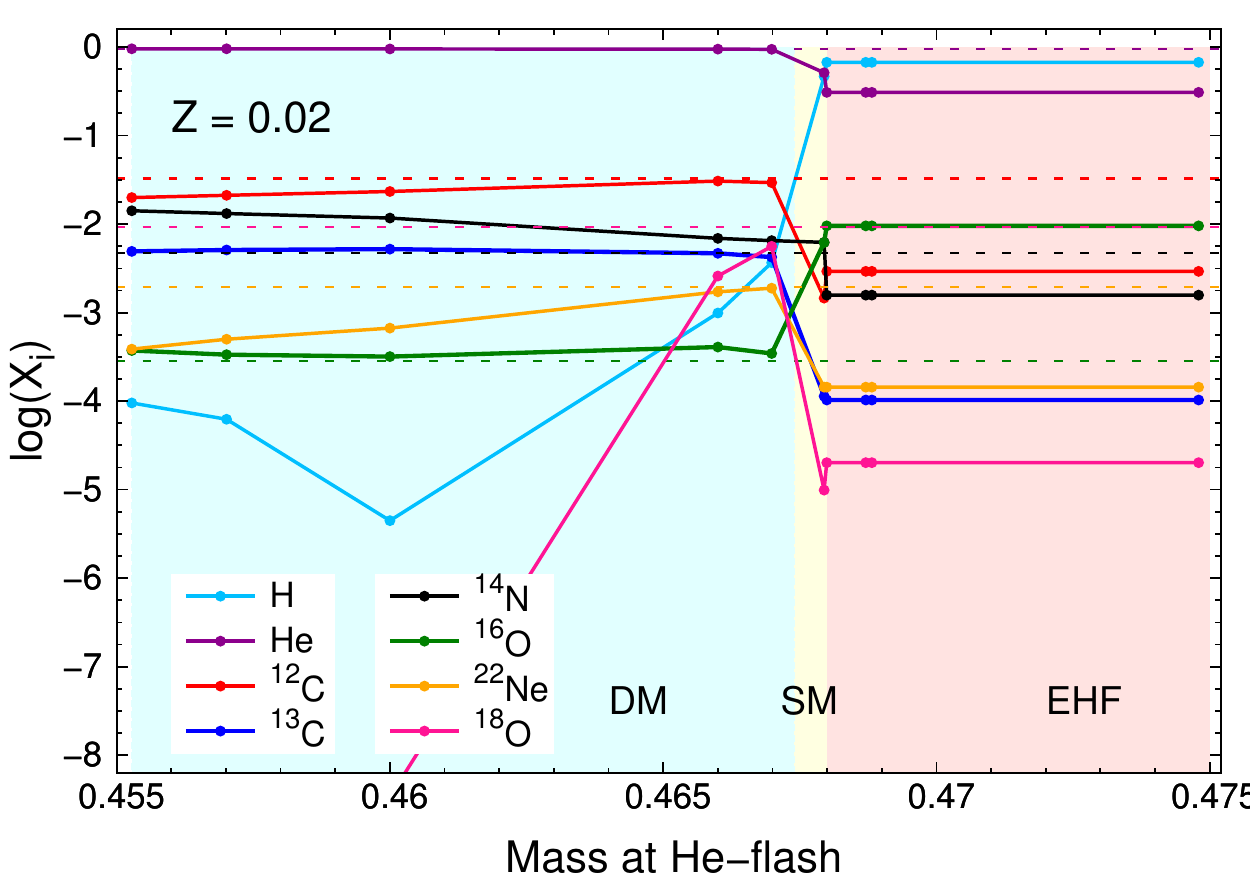}}
	\subfigure{\includegraphics[width=0.45\textwidth]{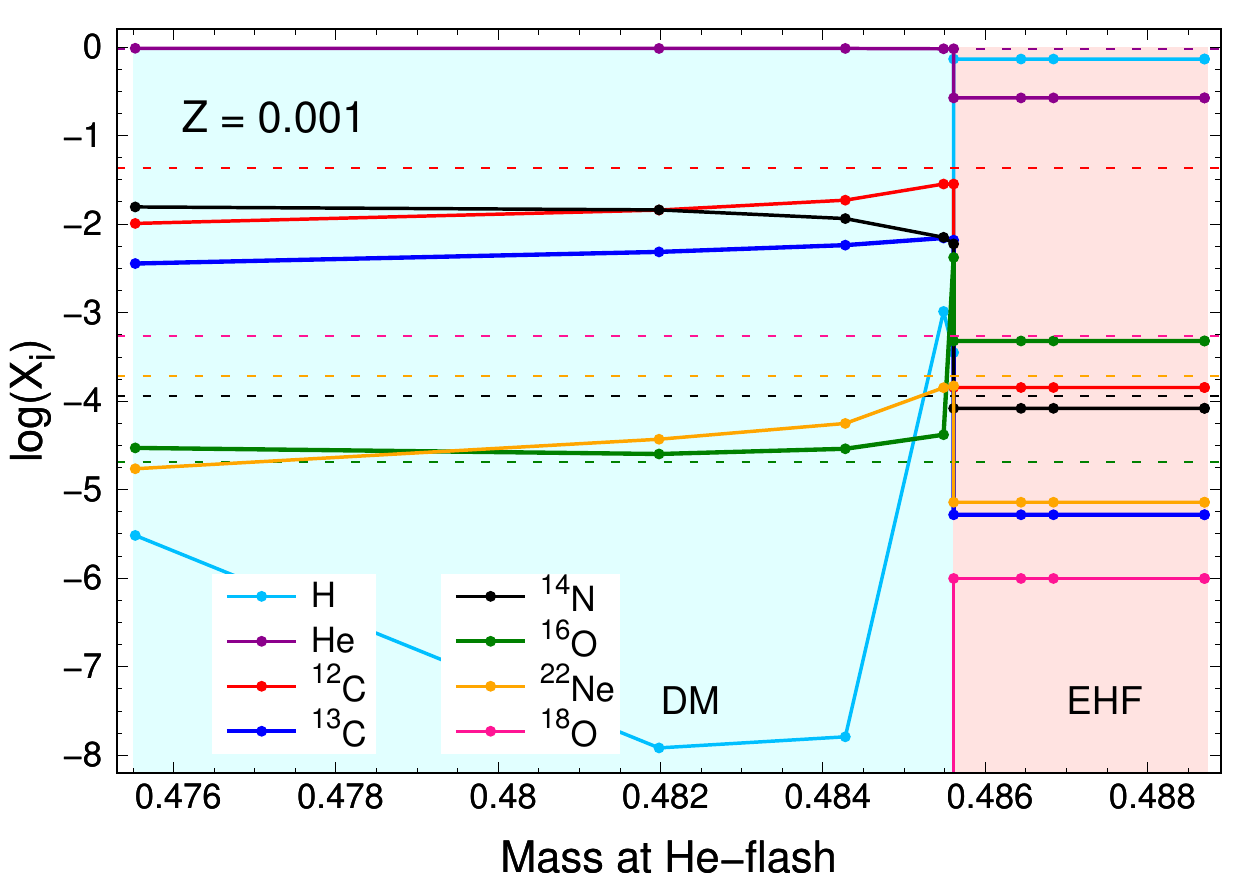}}
	\caption{Surface abundances by mass fraction at the onset of stable He burning in terms of the mass at He ignition (full lines). Are shown for comparison the typical abundances of the He-core after the primary He-flash has ended but before the He-subflashes (dashed lines). Upper panel corresponds to initial chemical compositions of ${\rm Z}=0.02$ and ${\rm Y}=0.285$. Lower panel corresponds to initial chemical compositions of ${\rm Z}=0.001$ and ${\rm Y}=0.247$.}
	\label{fig:abundances}
\end{figure}

\begin{table*}
\caption{Surface abundances of the late hot-flasher scenario at the ZAHB.}              
\label{table:1}      
\centering                                      
\begin{tabular}{l c c c c c c c c c}          
\hline\hline                        
Type$\!\!\!$ &	$^1$H	&		$^4$He	&		$^{12}$C	&			$^{13}$C	&			$^{14}$N	&			$^{16}$O	&	$^{18}$O	&		$^{20}$Ne	&			$^{22}$Ne \\
\hline  
 \multicolumn{10}{c}{ ${\rm Z}=0.02$, ${\rm Y}=0.285$ } \\
\texttt{SM}$\!\!$ &	$\!\!$0.4664 						&	$\!$0.5135$\!$ &	$\!$1.461$\times 10^{-3}\!$ &	$\!$1.135$\times 10^{-4}\!$ &	$\!$6.203$\times 10^{-3}\!$ &	$\!$6.178$\times 10^{-3}\!$ & $\!$9.891$\times 10^{-6}\!$	& $\!$1.809$\times 10^{-3}\!$ &	$\!$1.439$\times 10^{-4}$\\
\texttt{DM*}$\!\!$ &	$\!\!$3.659$\times 10^{-3}\!$ &	$\!$0.9426$\!$ &	$\!$2.933$\times 10^{-2}\!$ &	$\!$4.235$\times 10^{-3}\!$ &	$\!$6.515$\times 10^{-3}\!$ &	$\!$3.455$\times 10^{-4}\!$ & $\!$5.599$\times 10^{-3}\!$	& $\!$1.809$\times 10^{-3}\!$ &	$\!$1.891$\times 10^{-3}$\\
\texttt{DM*}$\!\!$ &	$\!\!$9.880$\times 10^{-4}\!$ &	$\!$0.9463$\!$ &	$\!$3.057$\times 10^{-2}\!$ &	$\!$4.677$\times 10^{-3}\!$ &	$\!$6.883$\times 10^{-3}\!$ &	$\!$4.081$\times 10^{-4}\!$ & $\!$2.593$\times 10^{-3}\!$	& $\!$1.809$\times 10^{-3}\!$ &	$\!$1.717$\times 10^{-3}$\\
\texttt{DM}$\!\!$ &	$\!\!$4.469$\times 10^{-6}\!$ &	$\!$0.9529$\!$ &	$\!$2.333$\times 10^{-2}\!$ &	$\!$5.212$\times 10^{-3}\!$ &	$\!$1.171$\times 10^{-2}\!$ &	$\!$3.188$\times 10^{-4}\!$ & $\!$3.372$\times 10^{-9}\!$	& $\!$1.809$\times 10^{-3}\!$ &	$\!$6.673$\times 10^{-4}$\\
\texttt{DM}$\!\!$ &	$\!\!$6.230$\times 10^{-5}\!$ &	$\!$0.9539$\!$ &	$\!$2.114$\times 10^{-2}\!$ &	$\!$5.096$\times 10^{-3}\!$ &	$\!$1.317$\times 10^{-2}\!$ &	$\!$3.342$\times 10^{-4}\!$ & $\!$4.346$\times 10^{-9}\!$	& $\!$1.809$\times 10^{-3}\!$ &	$\!$4.998$\times 10^{-4}$\\
\texttt{DM}$\!\!$ &	$\!\!$9.554$\times 10^{-5}\!$ &	$\!$0.9543$\!$ &	$\!$1.990$\times 10^{-2}\!$ &	$\!$4.918$\times 10^{-3}\!$ &	$\!$1.412$\times 10^{-2}\!$ &	$\!$3.724$\times 10^{-4}\!$ & $\!$5.094$\times 10^{-9}\!$	& $\!$1.809$\times 10^{-3}\!$ &	$\!$3.863$\times 10^{-4}$\\
\hline
  \multicolumn{10}{c}{ ${\rm Z}=0.001$ , ${\rm Y}=0.247$ } \\
\texttt{DM*}$\!\!$ & $\!\!$3.537$\times 10^{-4}\!$ &	$\!$0.9538$\!$ &	$\!$2.839$\times 10^{-2}\!$ & 	$\!$6.587$\times 10^{-3}\!$ & 	$\!$5.973$\times 10^{-3}\!$ & 	$\!$4.194$\times 10^{-3}\!$ & $\!$5.869$\times 10^{-9}\!$ & $\!$9.044$\times 10^{-5}\!$ & $\!$1.484$\times 10^{-4}$\\
\texttt{DM*}$\!\!$ & $\!\!$1.028$\times 10^{-3}\!$ &	$\!$0.9562$\!$ &	$\!$2.830$\times 10^{-2}\!$ & 	$\!$6.960$\times 10^{-3}\!$ & 	$\!$7.072$\times 10^{-3}\!$ & 	$\!$4.181$\times 10^{-5}\!$ & $\!$7.882$\times 10^{-8}\!$ & $\!$9.044$\times 10^{-5}\!$ & $\!$1.419$\times 10^{-4}$\\
\texttt{DM}$\!\!$ &	$\!\!$1.614$\times 10^{-8}\!$ &	$\!$0.9636$\!$ &	$\!$1.865$\times 10^{-2}\!$ & 	$\!$5.777$\times 10^{-3}\!$ & 	$\!$1.157$\times 10^{-2}\!$ & 	$\!$2.892$\times 10^{-5}\!$ & $\!$2.759$\times 10^{-11}\!\!$ & $\!$9.044$\times 10^{-5}\!$ & $\!$5.590$\times 10^{-5}$\\
\texttt{DM}$\!\!$ &	$\!\!$1.210$\times 10^{-8}\!$ &	$\!$0.9660$\!$ &	$\!$1.437$\times 10^{-2}\!$ &	$\!$4.844$\times 10^{-3}\!$ &	$\!$1.446$\times 10^{-2}\!$ &	$\!$2.527$\times 10^{-5}\!$ & $\!$2.284$\times 10^{-11}\!\!$ & $\!$9.044$\times 10^{-5}\!$ &	$\!$3.719$\times 10^{-5}$\\
\texttt{DM}$\!\!$ &	$\!\!$3.037$\times 10^{-6}\!$ &	$\!$0.9703$\!$ &	$\!$1.015$\times 10^{-2}\!$ &	$\!$3.587$\times 10^{-3}\!$ &	$\!$1.560$\times 10^{-2}\!$ &	$\!$2.963$\times 10^{-5}\!$ & $\!$1.471$\times 10^{-11}\!\!$ & $\!$9.044$\times 10^{-5}\!$ &	$\!$1.719$\times 10^{-5}$\\
\hline
  \multicolumn{10}{c}{ ${\rm Z}=0.02$, ${\rm Y}=0.4$  } \\
\texttt{SM}$\!\!$ &	$\!\!$0.2391$\!$ 			  &	$\!$0.7285 &	$\!$1.270$\times 10^{-2}\!$ & $\!$5.729$\times 10^{-5}\!$ &	$\!$6.877$\times 10^{-3}\!$ & $\!$0.3.944$\times 10^{-3}\!$ & $\!$2.464$\times 10^{-3}\!$ & $\!$1.809$\times 10^{-3}\!$ &	$\!$3.858$\times 10^{-4}\!$\\
\texttt{SM}$\!\!$ &	$\!\!$3.735$\times 10^{-2}\!$ &	$\!$0.9184 &	$\!$2.365$\times 10^{-2}\!$ & $\!$3.119$\times 10^{-6}\!$ &	$\!$8.870$\times 10^{-3}\!$ & $\!$0.5.117$\times 10^{-4}\!$ & $\!$4.719$\times 10^{-3}\!$ & $\!$1.809$\times 10^{-3}\!$ &	$\!$5.908$\times 10^{-4}\!$\\
\texttt{DM*}$\!\!$ & $\!\!$7.017$\times 10^{-3}\!$ & $\!$0.9444 &	$\!$2.268$\times 10^{-2}\!$ & $\!$4.931$\times 10^{-3}\!$ &	$\!$1.262$\times 10^{-2}\!$ & $\!$0.6.426$\times 10^{-4}\!$ & $\!$1.224$\times 10^{-3}\!$ & $\!$1.809$\times 10^{-3}\!$ &	$\!$5.849$\times 10^{-4}\!$\\
\texttt{DM*}$\!\!$ & $\!\!$2.582$\times 10^{-3}\!$ & $\!$0.9509 &	$\!$1.745$\times 10^{-2}\!$ & $\!$4.981$\times 10^{-3}\!$ &	$\!$1.703$\times 10^{-2}\!$ & $\!$0.6.557$\times 10^{-4}\!$ & $\!$6.761$\times 10^{-5}\!$ & $\!$1.809$\times 10^{-3}\!$ &	$\!$3.955$\times 10^{-4}\!$\\
\hline                                             
\end{tabular}
\end{table*}

Figure \ref{fig:abundances} depicts the surface abundances of different models at the ZAHB for different metallicities, in terms of the total mass of the model at He ignition. The mass ranges for the different flavours of hot flasher are shown in different colours. As in previous works, we find that the mass range for the different flavours of Hot-Flashers is dependent on the initial metallicity of the sequences. In fact, for ${\rm Z}=0.001$ we did not find a SM case, leaving the range for SM at  ${\rm Z}=0.001$ to be below $\delta {M_\star}^{\rm SM} < 10^{-6}\,M_{\odot}$. Instead, for ${\rm Z}=0.001$, we found a DM case where the maximum of energy liberation is much lower ($L/L_{\odot}\sim10^{8}$) than in standard DM events ($L/L_{\odot}\sim10^{10}$). In this case the mixing is still deep into the HeFCZ and the surface abundances change as in a DM case. This case is an intermediate case between a normal DM case and a SM* case where the depth of the outer convective mixing is much shallower. We label it as DM*. For ${\rm Z}=0.02$ we found all flavours of hot flasher including a DM* case. 

\begin{figure}
	\centering
	\includegraphics[width=0.48\textwidth,angle=0]{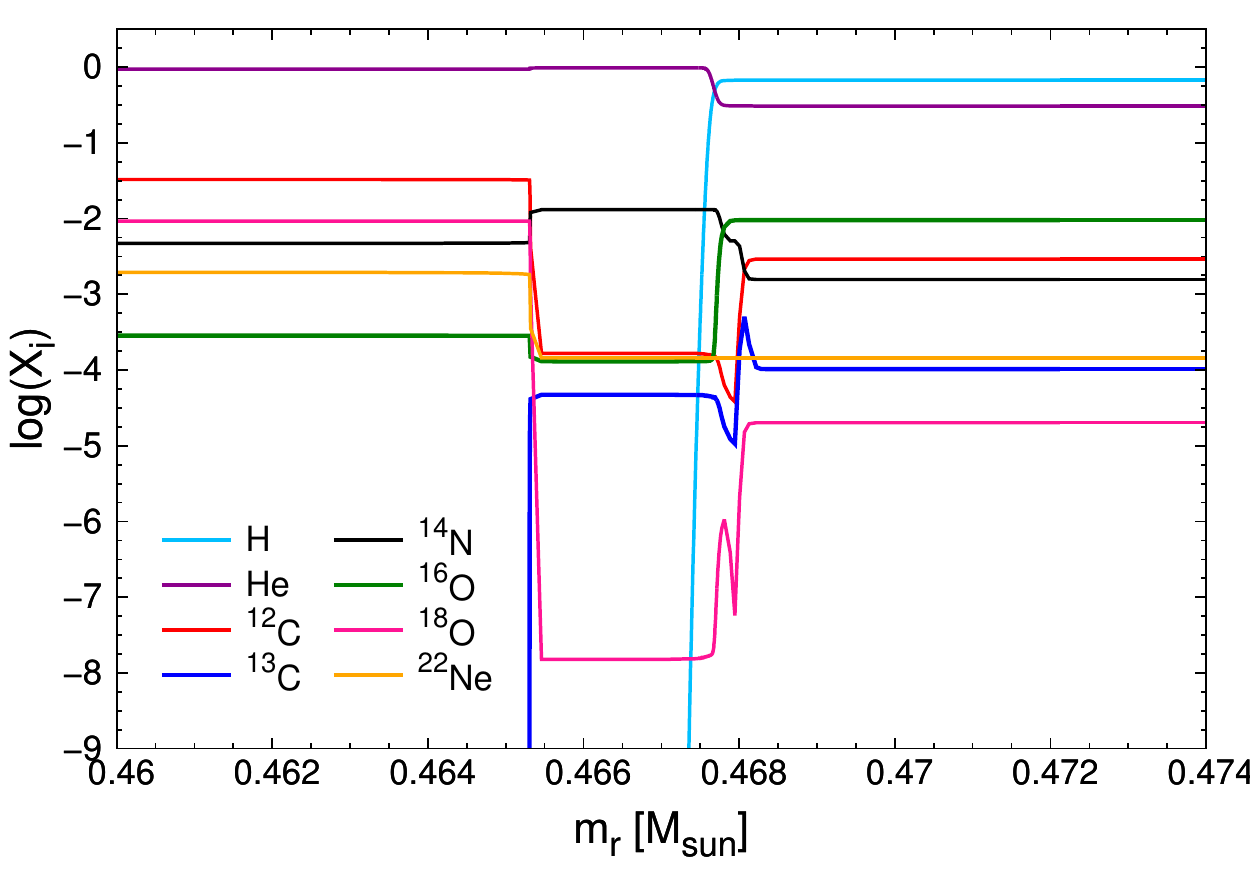}
	\caption{Typical chemical stratification below the H-rich envelope (at $m_r\gtrsim 0.468 M_\odot$) during the main He-core flash, for a model with initial chemical compositions of ${\rm Z}=0.02$ and ${\rm Y}=0.285$. The chemical inversions at about $m_r=0.468$ are a consequence of the H-shell burning.}
	\label{fig:idea}
\end{figure}

Now we discuss briefly the obtained surface abundances after the He-flash. In EHF cases the abundances are left unchanged and are the same as in the case where the He-flash occurs at the RGB tip. On the other hand, late hot-flashers lead to the dredging up of material from the stellar interior to the photosphere. Depending on the deepness of the mixing, different surface abundances arise. When shallow mixing (SM) occurs, the surface abundances of $^{12}$C and $^{16}$O decrease and the abundances of He and $^{14}$N increase respect to the EHF and canonical cases. This is because in SM events the envelope is mixed with the H-free region rich in He and $^{14}$N that lies  immediately below the H-rich envelope. This region displays the typical abundances left behind by the  stable CNO-burning during the RGB (see Fig. \ref{fig:idea}). Also, as H is diluted into the H-free material coming from the interior,  the surface H abundance  decreases. This is shown in Fig. \ref{fig:abundances}. The exact values will depend on metallicity and the depth of the outer convective zone. In deep mixing (DM) episodes, the H-rich envelope is mixed into the hot HeFCZ where large amounts of $^{12}$C have already been created by the 3$\alpha$-reaction during the primary He-core flash (see Fig. \ref{fig:idea}), and a violent H flash develops. As H is burned in a $^{12}$C rich environment, $^{13}$C is created. Once enough $^{13}$C has been created, H is burned with  $^{13}$C increasing the final $^{14}$N superficial abundance. In addition, during He flash the $^{14}$N (coming from previous CNO-burning on the RGB), is burned with $\alpha$ particles to produce $^{18}$O. In turn, $^{18}$O is burned producing $^{22}$Ne.
 Once the H-rich material is mixed and burned in the hot interior, the violent H-flash splits the HeFCZ into two (due to the huge energy release). The outer convective zone, driven by H-flash burning is too cold for He-burning to take place and $^{18}$O and $^{22}$Ne in the outer convective zone are left unchanged after that. As $^{18}$O and  $^{22}$Ne keep increasing during the He-flash, the exact final abundances of $^{18}$O and  $^{22}$Ne in DM-episodes depend on how many years after the peak of the He-flash does the mixing take place. As $^{12}$C also keeps increasing during the He-flash, meanwhile $^{14}$N keeps decreasing, the same holds for the abundances of these elements, with the difference that their exact values are also affected by H burning. All in all the $^{12}$C and $^{22}$superficial Ne abundances are higher for cases where the H flash takes more time to develop, meanwhile the $^{14}$N superficial abundance is lower for those cases. These are the DM cases closer to the SM region in Fig. \ref{fig:abundances}. 

The same holds for ${\rm Z}=0.001$ and for the ${\rm Y}=0.4$ sequences, with some quantitative differences due to the different initial compositions. The resulting surface abundances at the onset of stable core He burning for the case of late hot-flashers are listed in Table \ref{table:1}.

The typical differences between our derived surface abundances and the ones derived for \citet{2008A&A...491..253M} represent less than the 20\% of our surface abundances values for both the DM and SM cases. The exceptions are the abundances of C, H and O in the DM case. The larger difference is for the abundance of O, being the abundances derived by \citet{2008A&A...491..253M} around twice the values that we obtain. As \citet{2016arXiv160808739X} do not provide a classification of the flavour of the sequences in their tables (DM or SM), we compared those that are clearly deep mixing cases with our DM cases. The differences for He, C, N and O (the only surface abundances provided in \citealt{2016arXiv160808739X}) are always below 20\% except for the abundance of O, being their values also about twice the values of our derived O abundances. In all cases the differences are smaller than a factor of 2, being all abundances of the same order of magnitude. Therefore, the final surface abundances computed in this work for the different late hot-flasher scenarios are in good qualitative agreement with those of \citet{2016arXiv160808739X} (for ${\rm Z}=0.02$) and \citet{2008A&A...491..253M} (for ${\rm Z}=0.02$ and ${\rm Z}=0.001$).

\section{Stability analysis}
\label{sec:results}

\begin{figure*}[h!]
	\centering
\includegraphics[width=0.575\textwidth]{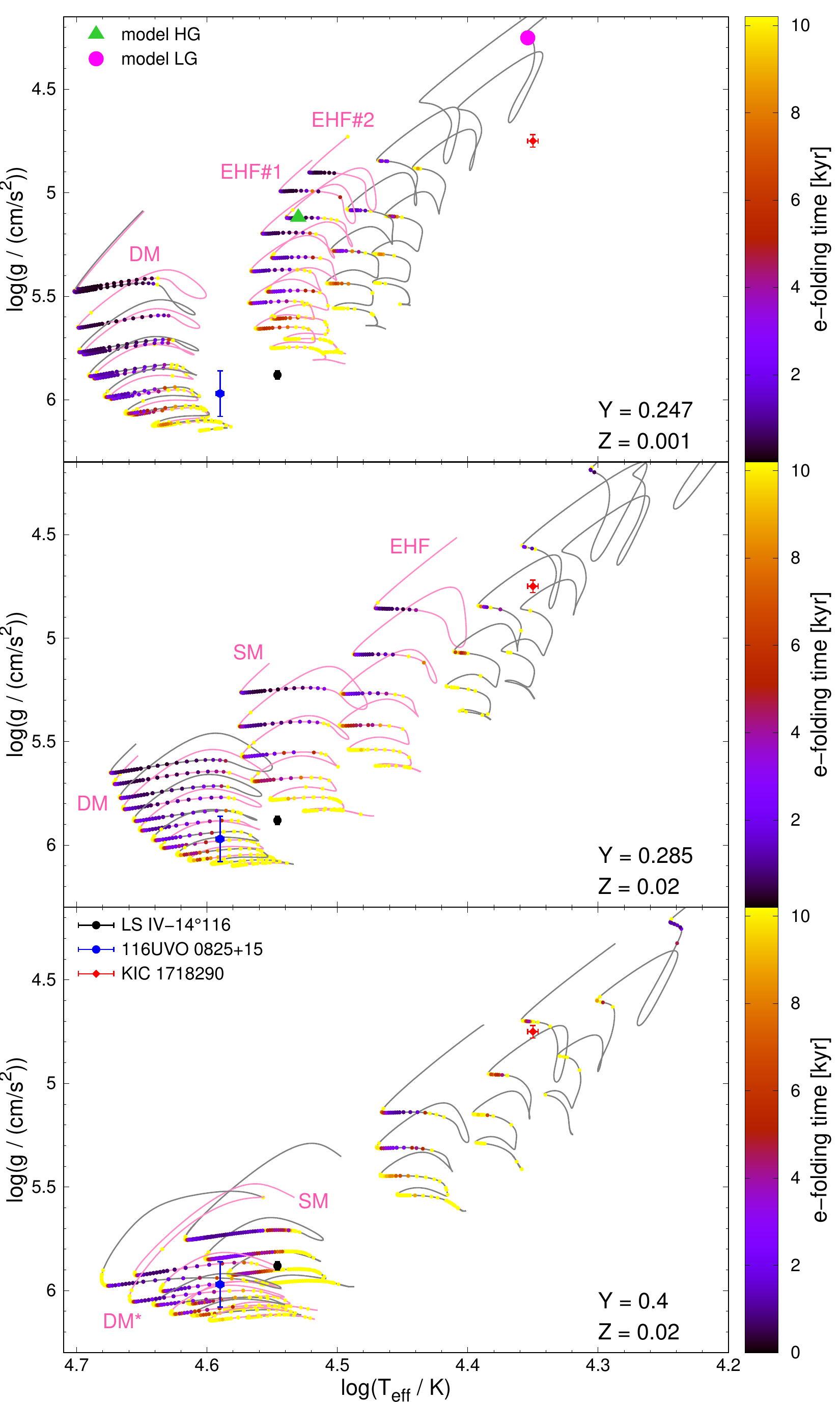}
	\caption{The phase of the He-subflashes of the evolutionary sequences in the $\log T_{\rm eff} - \log g$ diagram (grey and pink lines). Coloured points correspond to models with excited modes and the colour coding shows the minimum $e$-folding time for each model. Evolutionary sequences in pink lines have their pulsational quantities listed in Table \protect\ref{table:pul1}. Also shown for comparison is the location of the known He-rich hot subdwarf pulsators LS IV-14$^\circ$116 \protect\citep{2015A&A...576A..65R}, UVO 0825+15 \protect\citep{2017MNRAS.465.3101J}, and KIC 1718290 \protect\citep{2012ApJ...753L..17O}. The upper panel also shows the models discussed in Sect. \protect\ref{sec:driving}.}
	\label{fig:kiel}
\end{figure*}

\begin{figure*}
	\centering
	\includegraphics[width=0.7\textwidth]{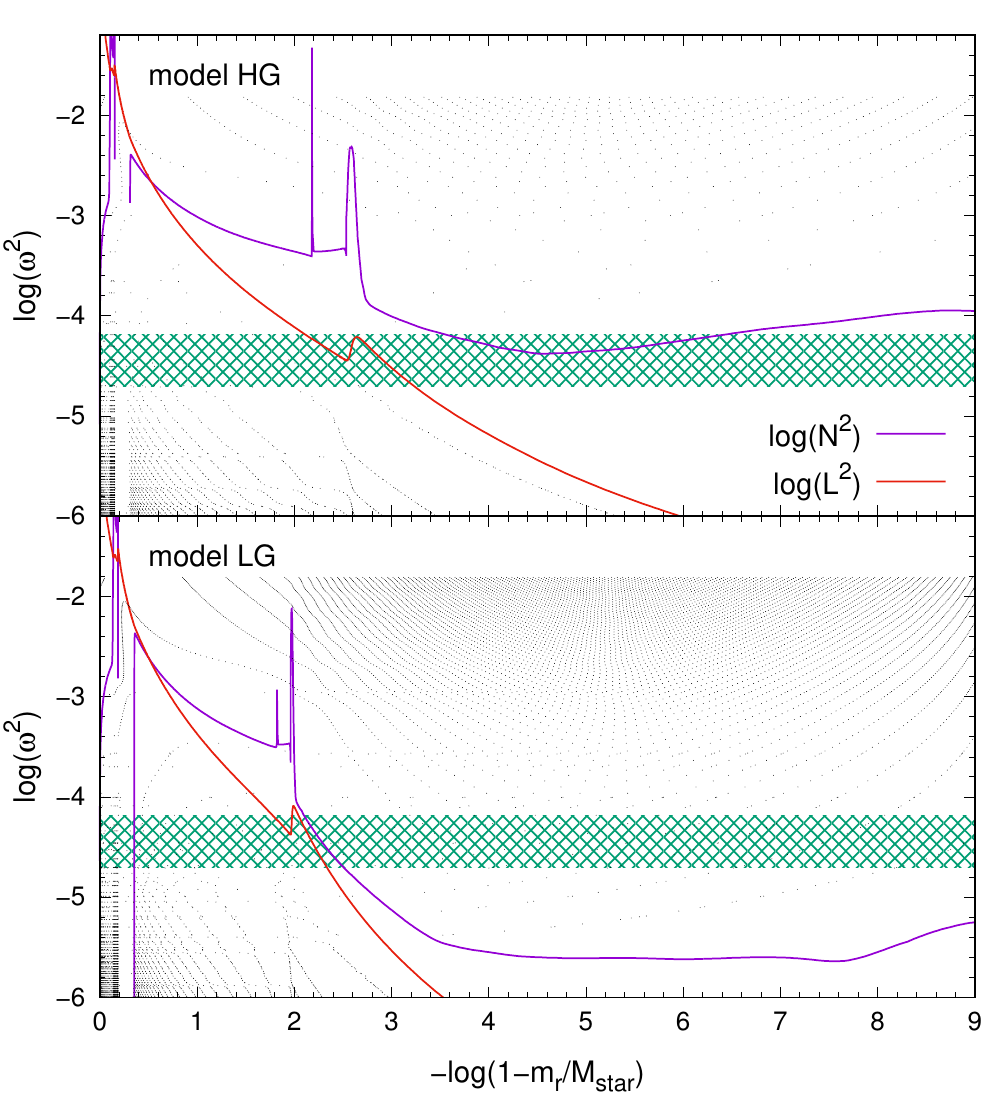}
	\caption{The logarithm of the squared Lamb and Brunt-Väisälä frequencies (violet and red lines) in terms of the outer mass fraction coordinate. Black points mark the nodes of the radial eigenfunctions. Upper panel corresponds to a model with $\log g=5.12$, $\log T_{\rm eff}/{\rm K}=
4.53$, $M_{\star}=0.4868\,M_{\star}$ (model HG) with excited modes. Lower panel correspond to a model with $\log g=4.25$, $\log T_{\rm eff}/{\rm K}=4.35$, $M_{\star}=0.4908\,M_{\odot}$ (model LG) and no excited modes. The region of periods in the range $\sim 880-1130\,$s is shown with a green strip. This correspond to the region of excited modes for model HG. At about $\log(1-m_r/M_{\star})=-0.2$ the Brunt-Väisälä frequency drops to zero in the convective zone. The pulsation driving region is located at the base of this convective zone.}
	\label{fig:frec} 
\end{figure*}

\begin{figure}
	\centering
	\includegraphics[width=0.49\textwidth]{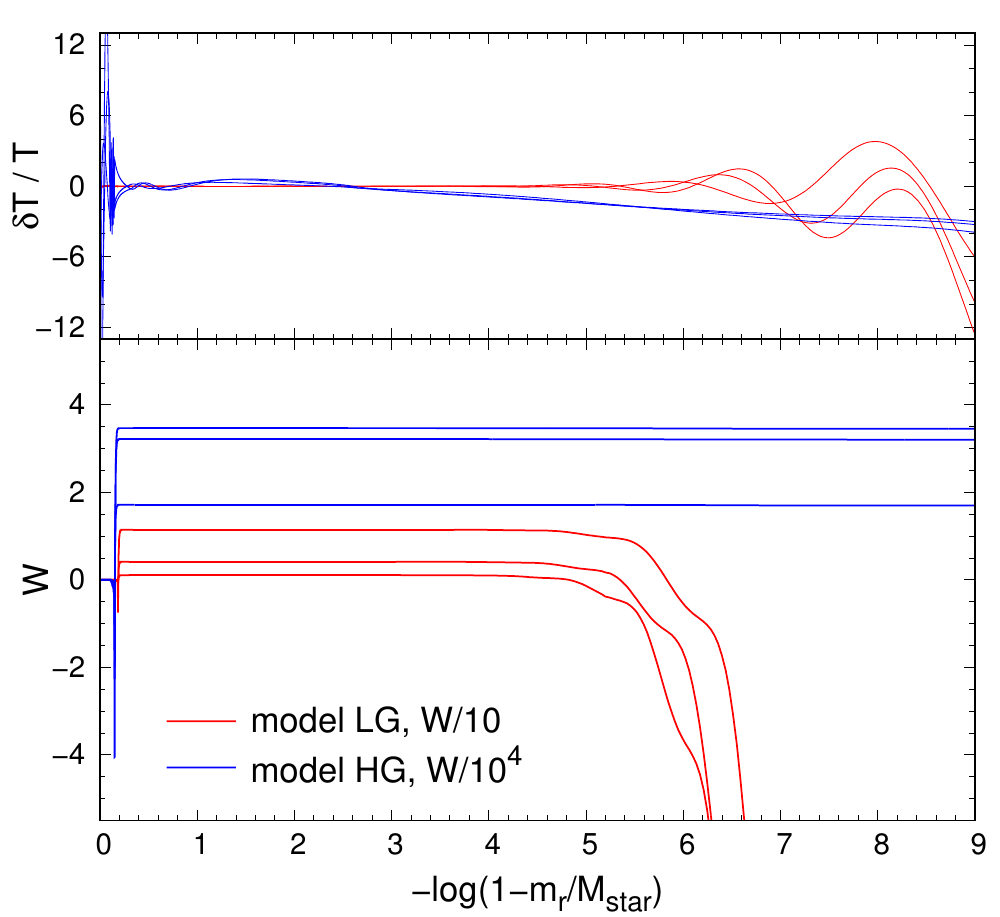}
	\caption{Upper panel: Lagrangian perturbation of the temperature for three modes of each of the models of Fig. \ref{fig:frec}. These modes have periods between 880-1130s and are excited by $\epsilon$ mechanism in the inner region ($-\log(1-m_r/M_{\star})\sim 0.2$). Blue (red) lines correspond to the modes of the model HG (LG). Lower panel: Running work integral for the same models. The quantities $W$ and $\delta T/T$ are normalized in order to have the radius perturbation at the surface $\delta r/R=1$. In the lower panel the running work integral is normalized arbitrarily for each model to be able to plot them in the same panel.}
	\label{fig:trabajo} 
\end{figure}

We performed a stability analysis over a total of 20 stellar evolutionary sequences for different values of initial chemical abundances and masses of the H-rich envelope. We computed non-adiabatic non-radial pulsations in stellar models going through He-subflashes. For each evolutionary sequence we performed a stability analysis over 300 to 800 stellar structure models (the exact value depending on initial chemical abundances), covering the whole subflashes stage. For each of these models we explored $\ell=1$ modes within a period range of 50 to 7000 s. For one of the evolutionary sequences  we also performed calculations of $\ell=2$ and $\ell=3$ modes, to determine the trend of the periods and excitation with the value of $\ell$. 

We found a new theoretical instability domain in the $\log T_{\rm eff} - \log g$ plane (also known as Kiel diagram). The domain of instability corresponds to low/intermediate-order $g$ modes excited by $\epsilon$ mechanism, in agreement of what was found for one evolutionary sequence by \citet{2011ApJ...741L...3M}. The period range of excited modes is from 200s to 2000s. We found that pulsations are only excited in the hottest models with $T_{\rm eff}\gtrsim 20000\,$K (see Fig. \ref{fig:kiel}). As a consequence pre-EHB models show $\epsilon$-mechanism driven pulsations during the He-subflashes, while pre-BHB models do not. Also, oscillations are excited in models with surface gravities higher than $\log g \simeq 4.8$. Within the hot-flasher scenario the He-subflashes occupy a region in the Kiel diagram of about 1 dex above the ZAHB, and never attain temperatures higher than $50000\,K$ (see Figs. \ref{fig:hr:DM} and \ref{fig:kiel}). The instability domain in the Kiel diagram predicted by our sequences is therefore the intersection of the locus of the evolutionary sequences during the He-subflashes and the temperature and gravity boundaries mentioned before ($T_{\rm eff}\gtrsim 20000\, K$ and $\log g \gtrsim 4.8$), see Fig. \ref{fig:kiel}. The location  of the ZAHB depends on both the initial metal (Z) and helium (Y) content of the models. In particular, in the He-enriched models the ZAHB is located at higher gravities due to the smaller He-cores and the consequent lower luminosities. While the high temperature and gravity boundaries of the instability domain are the direct consequence of the locus of the hot-flasher models in the Kiel diagram, the lower temperature and gravity boundaries requires a detailed analyses of the excitation and damping mechanisms.

\subsection{Driving and damping of pulsations}
\label{sec:driving}

All $g$ modes have relatively high amplitudes in the layers where the He-subflashes take place, and consequently undergo some degree of excitation through the $\epsilon$ mechanism. Yet, whether or not a mode is actually excited is determined by the competition between the driving mechanism and the damping of oscillations, mostly through
radiative damping. To understand the excitation and damping mechanisms it becomes useful to look at the mechanical structure of
the pre-HB models through a propagation diagram. In Fig. \ref{fig:frec} we show the propagation diagrams of two different
models during the peak of energy release in a He-subflash. The logarithm of the squared Brunt-Väisälä (Lamb) frequency is plotted with full violet (red) lines. There are two peaks of the Brunt-Väisälä frequency near $\log(1-m_r/M_{\star})=-2$. The outermost one corresponds to the He-H transition, and the innermost one, to the most external position reached by the outer edge of the convective zone during the main He flash, that reduces the abundance of He in 0.04\% and increases the abundance of $^{12}$C. The majority of the nodes of the radial eigenfunctions are clustered in the core, in particular in the regions of high values of
the Brunt-Väisälä frequency. At about $\log(1-m_r/M_{\star})=-0.2$, the Brunt-Väisälä frequency drops to zero in the convective zone (the zone without any nodes). The pulsation driving region is located at the base of this convective zone. The model on the upper panel of Fig. \ref{fig:frec} corresponds to a high-gravity, high-temperature model ($\log g=5.12$, $\log T_{\rm eff}/{\rm K}=4.53$ and $M_{\star}=0.4868\,M_{\star}$, hereafter model HG) that displays unstable modes with periods in the range $\sim 880-1130\,$s, while the lower panel of Fig. \ref{fig:frec} corresponds to a low-gravity, low-temperature model ($\log g=4.25$, $\log T_{\rm eff}/{\rm K}=4.35$ and $M_{\star}=0.4908\,M_{\odot}$, hereafter model LG, see also Fig. \ref{fig:kiel}) without unstable modes. The Brunt-Väisälä frequency ($N$) of model LG reaches lower values in the outer layers. $N$ can be written as\footnote{Where $P$, $\rho$, $\nabla_{\rm ad}$ and $\nabla$ are the local pressure, density, adiabatic and actual temperature gradient, $\chi_T=(\partial \ln P/\partial \ln T)_{\rho}$, $\chi_{\rho}=(\partial \ln P /\partial \ln \rho)_T$, $X_i$ is the abundance per mass of the specie $i$, $\chi_{X_i}=(\partial \ln P /\partial \ln X_i)_{\rho,T,\{X_{j\neq i}\}}$ and $n$ the total number of species considered.}\citep{1991ApJ...367..601B}
\begin{equation}
	N^2 = \frac{g^2 \rho}{P}\frac{\chi_T}{\chi_\rho}\,
\left(\nabla_{\rm ad}+\nabla -\frac{1}{\chi_T}
\sum_{i=1}^{n-1}\chi_{X_i}\frac{{\rm d}\ln X_i}{{\rm d}\ln P}\right).
\end{equation}
Therefore, $N$ is lower for lower values of the local gravity $g$. As a consequence, all the modes of model LG that have periods shorter than $3000\,$s ($\log\omega^2>-5.3$), oscillate as mixed modes ---i.e. they behave as $g$ modes in the core and as $p$-modes in the envelope, see e.g. \citet{2017A&ARv..25....1H}. Modes with periods in the range $\sim 880-1130\,$s are locally excited by the $\epsilon$ mechanism in both models. The running work integral\footnote{The running work integral at a radius coordinate $r$ represents the work done by the sphere of radius $r$ on the layer at radius $r$. The value of this function at the surface, $W(R)$, depends of the stability nature of modes. If $W(R)>0$ the mode is unstable. If $W(R)<0$ the mode is stable, see \citet{1989nos..book.....U}.} of three modes with periods in this range is shown for both models in the lower panel of Fig. \ref{fig:trabajo}. At $-\log(1-m_{r}/M_{\star})\sim0.2$ the work function  becomes $W>0$ due to the excitation by the $\epsilon$ mechanism. For model LG, the modes in the period range of $\sim 880-1130\,$s are mixed modes that oscillate as $p$-modes in the outer layers (lower panel of Fig. \ref{fig:frec}), where they have large amplitudes (upper panel of Fig. \ref{fig:trabajo}, red lines). As a consequence, these modes are strongly stabilized by radiative damping in the outer layers, and turn out to be globally stable modes. On the other hand, for model HG (upper panel of Fig. \ref{fig:frec}) the excited modes have frequencies at or below the local minimum of $N^2$ ($N^2_{\rm min}$; at around $-\log
(1-m_r/M_\star)\sim 4.5$). As a consequence, those modes oscillate as pure $g$ modes in the stellar core. On the contrary, modes
with frequencies higher than $N^2_{\rm min}$ oscillate as mixed modes. Due to their relative large amplitudes in the envelope,
modes with frequencies above $N^2_{\rm min}$ are also strongly stabilized by radiative damping in the outer layers. This explains the high frequency limit for the driving of pulsations. The low-frequency limit of the excited modes is a consequence of the radiative damping of pure $g$ modes in the core. This can be understood as follows. The local radiative damping rate $\gamma_{\rm rad}(r)$ depends on the local wavenumber of the modes. For modes with higher local wavenumbers, $\gamma_{\rm rad}(r)$ is higher. Locally, the temperature perturbations can be approximated as $\delta T/T\propto e^{-ik_r\,r}$, and within the diffusion approximation, $\gamma_{\rm rad}(r)$ for $g$
modes can be written as \citep{2013MNRAS.430.1736S,1998ApJ...493..412K}
\begin{equation}
	\gamma_{\rm rad}(r)=\frac{16\sigma T^3}{3\rho^2\kappa c_P}\, k_r^2,
	\label{eq:gama}
\end{equation}
where the temperature $T$, the density $\rho$, the opacity $\kappa$ and the specific heat $c_P$ are functions of the radial coordinate $r$, and $k_r$ is the local wavenumber. The dispersion relation for gravity modes with frequency $\omega<<N,L$ leads to
\begin{equation}
	k_r=\frac{\sqrt{\ell(\ell+1)}\,N}{r\,\omega}.
\label{eq:k}
\end{equation}
Lower frequencies have larger local wavenumber and are more strongly radiatively damped. For frequencies lower than a particular value, the damping in the radiative core became more important than the excitation by $\epsilon$ mechanism and the modes become stable.  

Once the excitation and damping mechanisms are understood, the origin of the low-gravity and low-temperature limits of the pulsation domain in the Kiel diagram (Fig. \ref{fig:kiel}), becomes clear. In models that undergo the He-subflashes at low surface gravities (like model LG) all modes with frequencies higher than $N_{\rm min}$ are mixed modes strongly damped by radiative diffusion in the outer regions. Also, due to the low value of $N_{\rm min}$, pure $g$ modes have too low frequencies and are strongly damped in the core. Consequently, no mode is actually excited in these models. The low-temperature limit of the pulsation domain is a consequence of the previous effect and the lower energy release of the later subflashes: as we move to lower effective temperatures, the first (and more intense) subflashes take place at too low gravities where modes are strongly radiatively damped in the outer layers of the star, and once the model has attained gravities closer to the ZAHB, the He-subflashes do not release enough energy to excite pulsations.

\subsection{Properties of excited modes}

\begin{figure}
	\centering
	\subfigure{\includegraphics[width=0.45\textwidth]{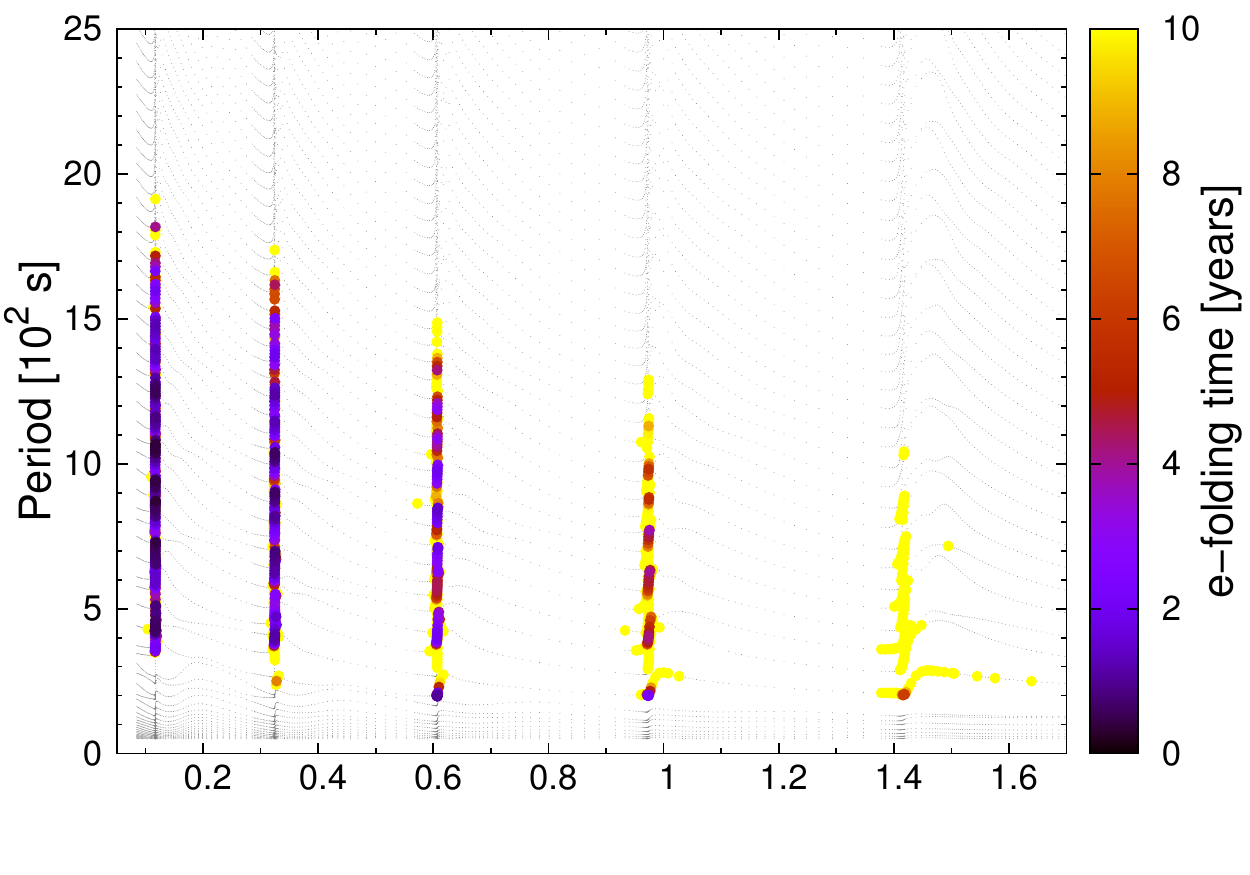}}
	\vskip-5ex
	\subfigure{\includegraphics[width=0.45\textwidth]{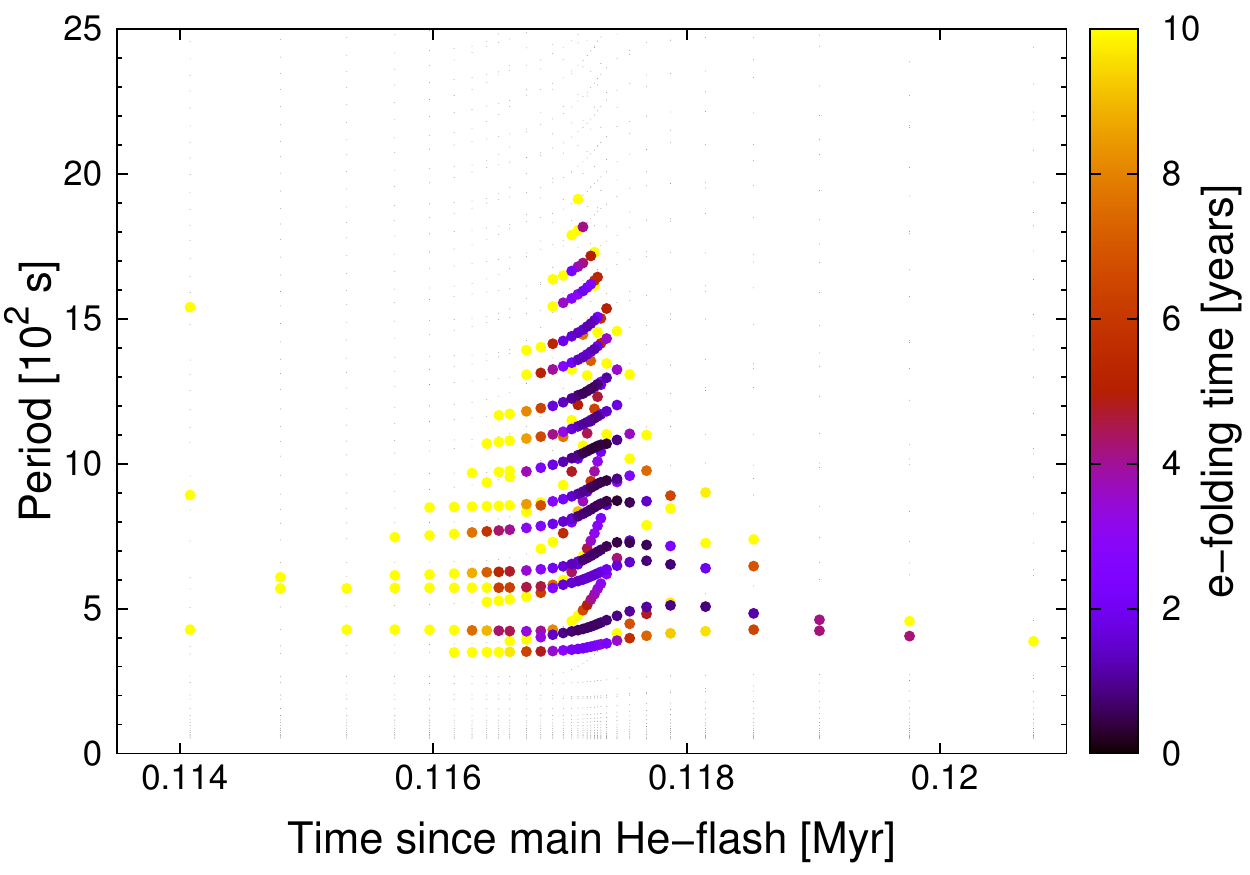}}   
	\caption{Upper panel: period vs. time of a sequence with ${\rm Z}=0.02$, ${\rm Y}=0.285$ and $M_{\star}=0.457\,M_{\odot}$. Excited periods are coloured with the $e$-folding time in colour coding. This sequence is marked as a pink DM in middle panel of Fig. \ref{fig:kiel}, and its pulsation properties are tabulated in Table \ref{table:pul1}. Lower panel: zoom on first subflash.}
	\label{fig:periods}
\end{figure}

\begin{figure}
	\centering
	\includegraphics[width=0.49\textwidth]{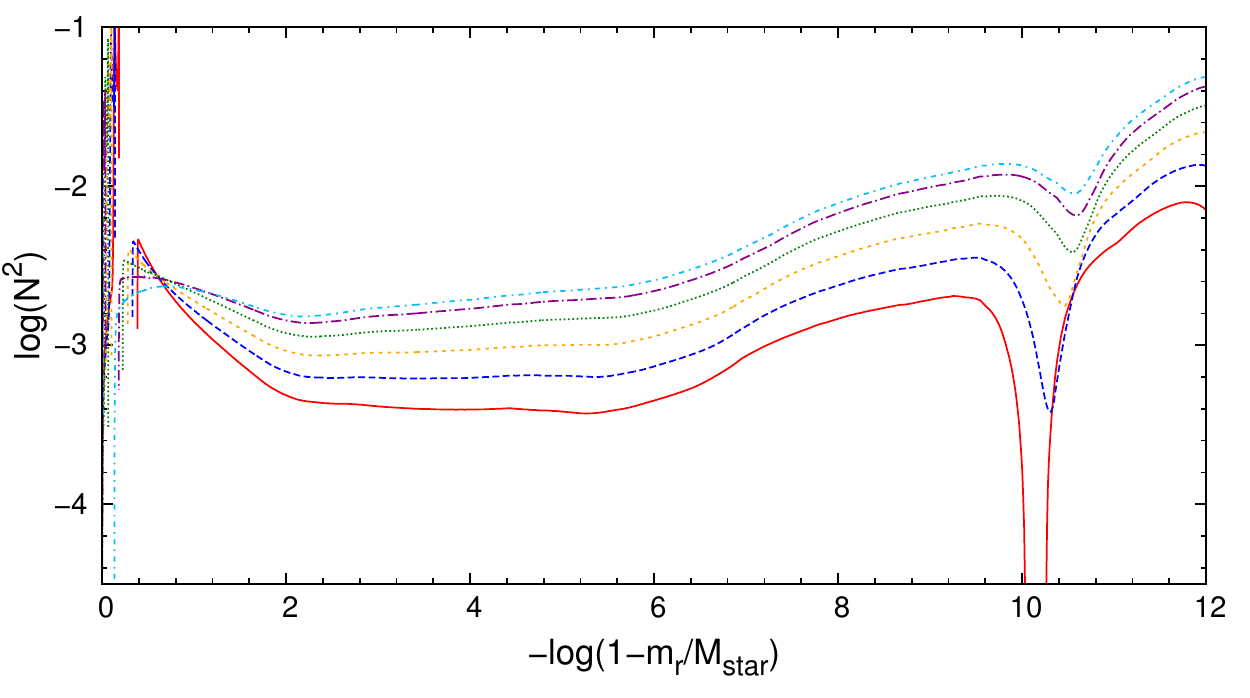}
	\caption{Brunt-Väisäla frequency ($N$) of models at the top of each subflash, for the same evolutionary sequence of Fig. \ref{fig:periods}. The red, full line corresponds to the first subflash, and upwards, to successive subflashes.}
	\label{fig:brunt} 
\end{figure}

We have obtained excited modes ($\ell=1$) with periods in the range $\sim 200-2000\,$s and radial orders in the range $\sim 1-25$. In all cases, the range  of excited periods becomes shorter with subsequent subflashes. This is exemplified in Fig. \ref{fig:periods} which shows the periods as a function of time for a sequence with ${\rm Z} = 0.02$, ${\rm Y} = 0.285$ and $M_{\star} = 0.457\,M_{\odot}$. The trend of the periods is a natural consequence of the continuous shrinking of the model as it approaches the ZAHB, which shifts the global pulsational properties to shorter periods. The short-period limit (high-frequency limit) is given by the local minimum of the Brunt-Väisälä frequency ($N^2_{\rm min}$) discussed in the previous section. $N^2_{\rm min}$ slightly increases in later subflashes, as shown in Fig. \ref{fig:brunt} which depicts the Brunt-Väisälä frequency for each subflash of the same evolutionary sequence of Fig. \ref{fig:periods}. As a consequence, the short-period limit slightly decreases in the last subflashes (see Fig. \ref{fig:periods}). This trend is more notorious for the long-period limit. This limit is due to the radiative damping of $g$ modes at the stellar core. For periods longer than a particular value, the radiative damping becomes more important than the excitation. For our models during the first subflash, for $\ell=1$, this happens for periods longer than about $\simeq 2000\,$s. As subflashes take place, the nuclear energy release is lower and the $\epsilon$ mechanism is less efficient. Therefore, the damping becomes more important than the excitation for shorter periods than in the first subflashes. 

\begin{figure}
	\centering
	\includegraphics[width=0.45\textwidth]{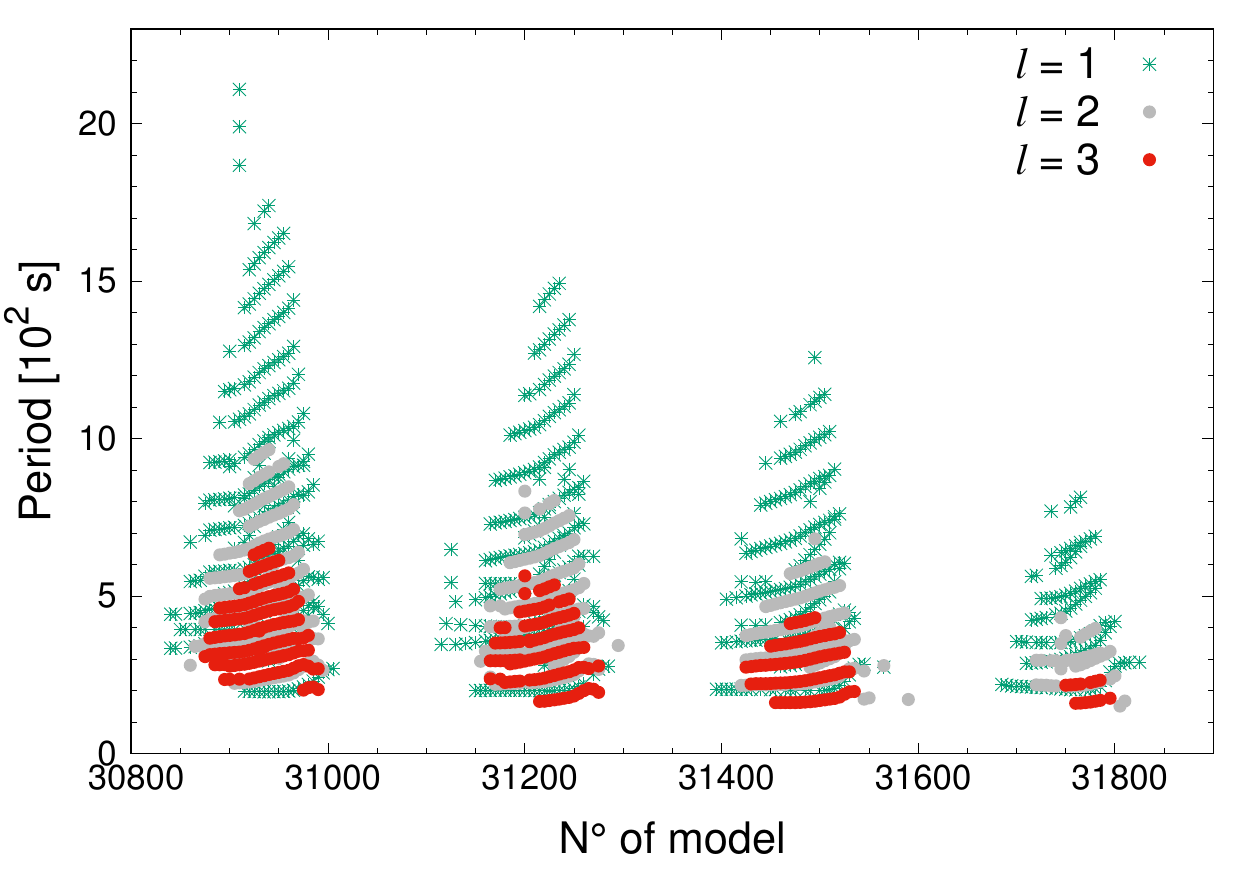}   
	\caption{Period vs. number of model of sequences with ${\rm Z}=0.02$, ${\rm Y}=0.4$ and ${M}_{\star}=0.445 {M}_{\odot}$, corresponding to $\ell=1$ (green points), $\ell=2$ (grey points) and $\ell=3$ (red points). This sequence is marked as Shallow Mixing case in Table \ref{table:pul1}.}
	\label{fig:navidad}
\end{figure}

The range of excited periods is also sensitive to the harmonic degree $\ell$. Meanwhile the short-period limit does not change with $\ell$ as it depends only on the value of $N$, the long-period limit of excited modes is lower for higher values of the harmonic degree $\ell$. This is illustrated in Fig. \ref{fig:navidad}, where we show the unstable periods for each computed value of $\ell$. This is a consequence of the dependence of the radiative damping on $\ell$ (see eqs. \ref{eq:gama} and \ref{eq:k}). As $\gamma_{\rm rad}$ increase with $\ell$, the period at which radiative damping overwhelms the excitation is shorter for higher values of the harmonic degree $\ell$.  

Another feature of the excitation by $\epsilon$ mechanism is the sharp transition between excited and non excited modes at the short-period limit. The sudden transition from a pure $g$-mode behaviour to a mixed mode one at $N^2_{\rm min}$ explains this sharp transition. On the contrary, the transition between excited and non-excited modes at the long-period limit occurs smoothly as the radiative damping increases with the increasing radial order.

Table \ref{table:pul1} shows some quantitative pulsational properties of selected sequences  in Fig. \ref{fig:kiel} at each subflash, before the sequences settle on the ZAHB. The Column 2 shows the maximum He-burning luminosity in each subflash. The Column 3 shows an estimate of the time span $\Delta t_{\rm ex}$ in which the modes are effectively being excited. Before and after this time interval the excitation due to $\epsilon$ mechanism is marginal. $\Delta t_{\rm ex}$ is assessed as the time interval in which modes are excited with $e$-folding times shorter than the duration of the subflash\footnote{As measured by the time spent with  He-burning luminosities $\log (L_{\rm He}/L_{\odot})>2$.}. The $e$-folding time ($\tau$) is a measure of the time required by a given unstable mode to grow to observable amplitudes. Modes with $\tau < \Delta t_{\rm ex}$ are excited by the $\epsilon$ mechanism long enough to grow to observable amplitudes. This allow us to estimate characteristic quantities for those modes that can actually reach observable amplitudes during the subflash. Therefore, all the other quantities in Table \ref{table:pul1} have been computed taking into account only unstable modes with $e$-folding times shorter than $\Delta t_{\rm ex}$. As shown in Column 4 of Table \ref{table:pul1} our computations predict that for all stellar models within the instability strip and for all the initial chemical compositions, there are several modes with values of $\tau$ lower than the time interval $\Delta t_{\rm ex}$. $\tau_{\rm min}$ and $<\tau>$ indicate the minimum and mean  $e$-folding times of all the significantly excited modes during each subflash. Note, in particular, that the shortest $e$-folding times ($\tau_{\rm min}$) are typically one order of magnitude lower than ($\Delta t_{\rm ex}$). As it is apparent from Table  \ref{table:pul1} (and also Figs. \ref{fig:kiel} and \ref{fig:periods}), the  $e$-folding times are shorter during the earlier subflashes, when the intensity of He-burning is higher and the $\epsilon$ mechanism is consequently more effective. The exception to this rule are those evolutionary sequences in which the first subflashes occur at lower surface gravities than the limit of the instability strip.

Another interesting property predicted by our computations is the rate of period change $\dot{P}$ of the  unstable modes. Due to the large structural changes driven by the sudden energy injection in the He-subflashes, the periods of the normal $g$ modes are strongly affected with typical values of $\dot{P}$ in the range of $\langle\dot{P}\rangle \sim 10^{-5}-10^{-7}\,$s/s (see Table \ref{table:pul1}). This corresponds to a period drift typically between 1 and 300 seconds per year that could be easily measured. These values are even higher than the typical $\langle\dot{P}\rangle$ derived by \citet{2008A&A...489.1201S} for radial modes of RR-Lyrae at the subflashes phase, $\langle\dot{P}\rangle \sim 10^{-8}\,$s/s. This is not surprising since, contrary to the RR-Lyrae case, the modes excited in our compact models are pure $g$ modes. This kind of modes lives in the core and are evanescent in the convective region that develops due to the He-flash. Therefore, pure $g$ modes are very sensitive to changes in the structure of the stellar core.

\begin{table*}
\caption{Pulsational properties of the selected sequences in Fig. \ref{fig:kiel}.  }              
\label{table:pul1}      
\centering   
\begin{tabular}{ccccccccccc}
\hline\hline
$\!\!$\#Subflash $\!\!$& $\!\!\!$$\log(L_{\rm He}^{\rm max}\!/L_{\odot})$ $\!\!\!\!\!$& $\Delta t_{\rm ex}$ & $\!\!$\#modes with$\!\!\!$ & $\langle\tau\rangle$ & $\tau_{\rm min}$ & $P_{\rm max}$ & $P_{\rm min}$ & $\langle\dot{P}\rangle$ & $\sigma_{\dot P}$ & $\dot P_{\rm max}$ \\
 & & [yr] & $\tau<\Delta t_{\rm ex}$ & [yr] & [yr] & [s] & [s] & [s/s] & [s/s] & [s/s] \\
\hline
 \multicolumn{10}{c}{ Deep Mixing (DM), ${\rm Z}=0.02$ and ${\rm Y}=0.285$  }\\
1 & 4.3 & 3454 & 162 & 1.46E+03 & 3.30E+02 & 1665.5 & 355.8 & 1.03E-05 & 1.51E-05 & 8.63E-05 \\
2 & 4.0 & 4809 & 114 & 2.23E+03 & 6.24E+02 & 1618.2 & 374.9 & 3.35E-06 & 4.94E-06 & 2.94E-05 \\
3 & 3.6 & 9489 & 111 & 4.30E+03 & 1.02E+03 & 1364.9 & 198.0 & 1.72E-06 & 2.44E-06 & 1.15E-05 \\
4 & 3.2 & 12375 & 69 & 6.85E+03 & 2.21E+03 & 1130.5 & 199.8 & 9.55E-07 & 1.03E-06 & 3.59E-06 \\
5 & 2.7 & 19050 & 34 & 1.29E+04 & 6.22E+03 & 732.5 & 202.3 & 4.53E-07 & 4.34E-07 & 1.62E-06 \\
 \multicolumn{10}{c}{ Shallow Mixing (SM), ${\rm Z}=0.02$ and ${\rm Y}=0.285$ }\\
1 & 4.3 & 3454 & 123 & 1.46E+03 & 3.30E+02 & 1665.5 & 355.8 & 1.03E-05 & 1.51E-05 & 8.63E-05 \\
2 & 4.0 & 4809 & 92 & 2.23E+03 & 6.24E+02 & 1618.2 & 374.9 & 3.35E-06 & 4.94E-06 & 2.94E-05 \\
3 & 3.6 & 9489 & 66 & 4.30E+03 & 1.02E+03 & 1364.9 & 198.0 & 1.72E-06 & 2.44E-06 & 1.15E-05 \\
4 & 3.2 & 12375 & 51 & 6.85E+03 & 2.21E+03 & 1130.5 & 199.8 & 9.55E-07 & 1.03E-06 & 3.59E-06 \\
5 & 2.7 & 19050 & 11 & 1.29E+04 & 6.22E+03 & 732.5 & 202.3 & 4.53E-07 & 4.34E-07 & 1.62E-06 \\
\multicolumn{10}{c}{ Early Hot Flasher (EHF) ${\rm Z}=0.02$ and ${\rm Y}=0.285$ }\\
1 & 4.4 & 904 & 15 & 6.63E+02 & 3.33E+02 & 1287.8 & 874.5 & 4.23E-05 & 3.33E-05 & 1.38E-04 \\
2 & 4.0 & 6061 & 62 & 2.36E+03 & 7.45E+02 & 1494.0 & 643.3 & 1.54E-05 & 1.27E-05 & 6.15E-05 \\
3 & 3.6 & 5197 & 41 & 3.00E+03 & 1.31E+03 & 1223.1 & 423.2 & 5.71E-06 & 4.40E-06 & 2.10E-05 \\
4 & 3.2 & 9659 & 30 & 6.56E+03 & 3.94E+03 & 997.7 & 435.0 & 2.18E-06 & 1.63E-06 & 5.75E-06 \\
5 & 2.8 & 14499 & 14 & 1.22E+04 & 9.36E+03 & 605.5 & 379.3 & 9.15E-07 & 6.17E-07 & 2.41E-06 \\
\hline
 \multicolumn{10}{c}{ Deep Mixing (DM), ${\rm Z}=0.001$ and ${\rm Y}=0.247$ }\\
1 & 4.7 & 1171 & 125 & 5.23E+02 & 9.86E+01 & 1688.0 & 417.1 & 5.29E-05 & 5.63E-05 & 3.11E-04 \\
2 & 4.4 & 2953 & 165 & 1.31E+03 & 1.60E+02 & 1847.0 & 366.7 & 1.68E-05 & 2.05E-05 & 9.73E-05 \\
3 & 4.1 & 4791 & 124 & 2.19E+03 & 4.74E+02 & 2482.9 & 383.6 & 5.81E-06 & 7.52E-06 & 3.92E-05 \\
4 & 3.8 & 7336 & 89 & 3.46E+03 & 1.11E+03 & 1378.4 & 204.5 & 2.43E-06 & 3.01E-06 & 1.41E-05 \\
5 & 3.4 & 11888 & 76 & 5.54E+03 & 1.55E+03 & 1157.6 & 202.3 & 1.50E-06 & 1.66E-06 & 6.17E-06 \\
6 & 3.0 & 20884 & 63 & 1.03E+04 & 3.72E+03 & 931.3 & 205.9 & 1.43E-06 & 2.98E-06 & 2.17E-05 \\
7 & 2.5 & 16504 & 5 & 1.35E+04 & 1.14E+04 & 220.4 & 211.7 & 3.03E-07 & 2.27E-07 & 7.15E-07 \\
\multicolumn{10}{c}{ Early Hot Flasher (EHF\#1), ${\rm Z}=0.001$ and ${\rm Y}=0.247$}\\
1 & 4.5 & 2176 & 54 & 1.03E+03 & 2.36E+02 & 1492.8 & 900.9 & 5.36E-05 & 3.87E-05 & 1.32E-04 \\
2 & 4.1 & 1948 & 45 & 1.03E+03 & 4.74E+02 & 1277.7 & 637.7 & 2.27E-05 & 1.46E-05 & 6.30E-05 \\
3 & 3.8 & 3249 & 34 & 1.84E+03 & 1.05E+03 & 1226.7 & 617.4 & 9.14E-06 & 6.21E-06 & 2.71E-05 \\
4 & 3.5 & 12853 & 53 & 5.53E+03 & 1.62E+03 & 1320.8 & 395.9 & 3.70E-06 & 2.39E-06 & 9.77E-06 \\
5 & 3.1 & 23348 & 44 & 1.11E+04 & 5.51E+03 & 1680.4 & 388.8 & 1.76E-06 & 1.01E-06 & 4.03E-06 \\
\multicolumn{10}{c}{ Early Hot Flasher (EHF\#2), ${\rm Z}=0.001$ and ${\rm Y}=0.247$ }\\
1 & 4.5 & 534 & 17 & 4.01E+02 & 2.27E+02 & 1268.1 & 832.4 & 4.81E-05 & 3.88E-05 & 1.33E-04 \\
2 & 4.1 & 2249 & 42 & 1.17E+03 & 4.57E+02 & 1530.3 & 637.9 & 1.94E-05 & 1.78E-05 & 5.74E-05 \\
3 & 3.8 & 3425 & 33 & 1.97E+03 & 1.16E+03 & 1222.8 & 418.9 & 8.06E-06 & 8.81E-06 & 2.64E-05 \\
4 & 3.5 & 7773 & 45 & 4.42E+03 & 2.39E+03 & 1160.6 & 409.5 & 4.01E-06 & 4.88E-06 & 9.78E-06 \\
5 & 3.0 & 10318 & 25 & 7.40E+03 & 5.63E+03 & 782.3 & 392.4 & 2.64E-06 & 3.92E-06 & 1.27E-05 \\
6 & 2.6 & 18704 & 5 & 1.56E+04 & 1.22E+04 & 1098.1 & 397.6 & 6.35E-07 & 1.06E-06 & 2.51E-06 \\
\hline
 \multicolumn{10}{c}{ Deep Mixing (DM*), ${\rm Z}=0.02$ and ${\rm Y}=0.4$  }\\
1 & 3.6 & 10361 & 116 & 4.68E+03 & 1.01E+03 & 1443.3 & 193.9 & 8.30E-07 & 1.48E-06 & 7.34E-06 \\
2 & 3.1 & 14353 & 67 & 8.71E+03 & 2.84E+03 & 1105.1 & 196.0 & 2.37E-07 & 4.09E-07 & 1.65E-06 \\
3 & 2.5 & 21340 & 17 & 1.63E+04 & 9.47E+03 & 679.7 & 198.0 & 1.64E-07 & 1.99E-07 & 7.21E-07 \\
 \multicolumn{10}{c}{ Shallow Mixing (SM), ${\rm Z}=0.02$ and ${\rm Y}=0.4$ }\\
1 & 3.9 & 7390 & 147 & 3.21E+03 & 6.14E+02 & 1606.7 & 195.9 & 1.74E-06 & 2.81E-06 & 1.68E-05 \\
2 & 3.4 & 11049 & 91 & 5.53E+03 & 1.45E+03 & 1345.2 & 198.9 & 1.03E-06 & 1.42E-06 & 6.22E-06 \\
3 & 3.0 & 18389 & 62 & 1.03E+04 & 3.75E+03 & 1012.4 & 199.6 & 5.42E-07 & 6.31E-07 & 2.49E-06 \\
4 & 2.4 & 14663 & 3 & 1.37E+04 & 1.34E+04 & 203.9 & 202.1 & 2.04E-07 & 1.83E-07 & 5.74E-07 \\
\hline
\end{tabular}
\tablefoot{In order the columns are: the number of the subflash, the maximum luminosity due to nuclear energy liberation in each subflash ($L_{\rm He}^{\rm max}$), the time span in which modes are effectively being excited ($\Delta t_{\rm ex}$, see the text for an explanation), the number of modes with $e$-folding times shorter than $\Delta t_{\rm ex}$, the mean and minimum $e$-folding times ($\langle\tau\rangle$, $\tau_{\rm min}$), the maximum and minimum excited periods ($P_{\rm max}$, $P_{\rm min}$), the mean period change rate ($\langle\dot P\rangle$), the standard deviation of the period change rates for each subflash ($\sigma_{\dot P}$), and the maximum value of the period change rates ($\dot P_{\rm max}$). The evolutionary sequences for which these quantities are tabulated are highlighted in Fig. \ref{fig:kiel}.}
\end{table*}

\section{Discussion}
\label{sec:disc}
 We now compare our predictions with the pulsational properties of the small number of known pulsating He-rich subdwarfs with mild He-enhancements, namely LS IV-14$^\circ$116 with observed periods $P\simeq 1950-5080\,$s ($T_{\rm eff}=35150\pm111$, $\log g=5.88\pm 0.02$, $\log n(He)/n(H)=-0.62\pm 0.01$, \citealt{2015A&A...576A..65R,2005A&A...437L..51A,2011ApJ...734...59G}), KIC 1718290 with $P\simeq 1-12$h ($T_{\rm eff}=22100$, $\log g=4.72$, $\log n(He)/n(H)=-0.45$, \citealt{2012ApJ...753L..17O}) and UVO 0825+156 with $P\simeq 10.8-13.3$h ($T_{\rm eff}=38900\pm270$, $\log g=5.97\pm 0.11$,$ \log n(He)/n(H)=-0.57\pm 0.01$,
  \citealt{2017MNRAS.465.3101J}) ---see Fig. \ref{fig:kiel}. Interestingly, as shown in Fig. \ref{fig:kiel}, all these stars fall within the range of surface gravities and temperatures predicted by our computations of the $\epsilon$ mechanism acting on hot-flasher sequences. The fact that the $\epsilon$ mechanism qualitatively predicts the excitation of $g$ modes in stellar models with similar values of $T_{\rm eff}$, $\log g$, and $\log n(He)/n(H)$ is very encouraging. Quantitatively, however, the range of excited periods in our sequences ($P \sim 200-2000\,$s) can only account for the shortest observed periods in these stars. This could be pointing to shortcomings in the stellar evolution models in this very fast and badly tested stage of the evolution. If He-subflashes were more intense than predicted by our sequences, the range of excited periods would reach longer periods. This is particularly interesting in the case of LS  IV-14$^\circ$116 which shows periods below $2\,$h and whose $T_{\rm eff}$, $\log g$, and $\log n(He)/n(H)$ values are well reproduced by our He-enhanced sequences ---bottom panel, Fig. \ref{fig:kiel}. On the other hand, due to the very strong radiative damping of higher-order $g$ modes, the longest  periods of the order of 10 h shown by KIC 1718290 and UVO 0825+156 can hardly be explained by the $\epsilon$ mechanism. 

Another interesting question concerns the number of pre-HB stars expected to pulsate due to the $\epsilon$ mechanism.
From Table \ref{table:pul1} we see that, within the hot-flasher scenario, the total amount of time spent by the sequences in the stages of pulsation instability is between $\sim 35000\,$yr and $\sim 65000\,$yr. Therefore, comparing  this time with the timescale typical of the whole pre-HB ($\sim 2\,$Myr, see Fig. \ref{fig:lum:time}) we can estimate that $\sim 2-3\,$\% of all pre-EHB stars should be pulsating due to the $\epsilon$ mechanism. The question now turns to how many stars in the current sdB samples \citep{2014ASPC..481...83F,2017A&A...600A..50G} are in the pre-EHB phase. Due to the short duration of this phase ($\sim 2\,$Myr) in comparison with the core He-burning phase ($\sim 100\,$Myr) pre-EHB stars should be 50 times less common than quiescent He-burning EHB stars in a volume-limited sample. However, pre-EHB stars can be up to one order of magnitude more luminous than EHB stars and can consequently be detected in  much larger volumes (up to $30$ times larger) for magnitude-limited samples. Consequently, the number of pre-EHB stars in a sample of field subdwarfs could be much higher than just a small percentage of the total. Under the optimistic assumption that most He-rich subdwarfs are indeed Hot-flashers on the pre-HB, we should expect $\sim 1$  out of the $\sim 50$ field He-rich subdwarfs in the sample of \cite{2014ASPC..481...83F}   to be pulsating by the  $\epsilon$ mechanism ---or $\sim 10$ out of the $\sim 500$ He-rich subdwarfs in the sample of \cite{2017A&A...600A..50G}. But even if we make the conservative assumption that only  a fraction of $\sim 0.0005$\footnote{Which is the ratio between the duration of the $\epsilon$-mechanism unstable phase ($\sim 50000\,$yr) to the typical length of the whole He-core burning EHB phase ($\sim100\,$Myr).} of the whole sdB sample is undergoing  $\epsilon$-mechanism driven  pulsations, we still find that at least $\sim 3$ stars should be pulsating by the $\epsilon$ mechanism in the sample of 5613 sdB stars presented by \cite{2017A&A...600A..50G}.

Pulsating subdwarfs have been observed in globular clusters like $\omega$ Cen \citep{2016A&A...589A...1R} and NGC 2808 \citep{2013ApJ...777L..22B} and one might wonder about the possibility of detecting $\epsilon$-mechanism driven pulsations in those clusters. The most favourable case is $\omega$ Cen, which is the most massive globular cluster and, in addition, has ~30\% of its horizontal branch stars in the blue hook \citep{2015Natur.523..318T}. A simple estimate of the expected number of pre-horizontal branch pulsating stars can be done by means of Renzini’s evolutionary flux method \citep{2011spug.book.....G}. Within this approximation, the number of stars ($N_j$) in a given stellar population (with total brightness $L_t$) for an evolutionary phase of duration $t_j$ is,
\begin{equation}
	N_j=B(t)\,L_t\,t_j
\end{equation}
where $B(t)$ is the specific evolutionary flux, that is very close to $2\times10^{-11}$ stars yr$^{-1} L_{\odot}^{-1}$ \citep{2011spug.book.....G}. In the case of $\omega$ Cen ($\log L/L_{\odot}=6.2$; \citealt{2010arXiv1012.3224H,1994ApJS...95..107W}), taking $\sim50000\,$yr for the total duration of the excitation of $\epsilon$-driven pulsations and that only 30\% of $\omega$ Cen horizontal branch stars are in the blue hook, we get $N_j\simeq 0.4$. Consequently, about only one star in the whole cluster is expected to be in this stage. The situation is worse in other clusters, with the total number of EHB stars in Globular Clusters well below 1000. Therefore, based on these simple considerations it seems unlikely to find stars pulsating by $\epsilon$ mechanism in the EHB of Globular Clusters.

\section{Summary and conclusions}
\label{sec:conclusion}
We performed non-adiabatic computations of stellar pulsations in pre-hot Horizontal Branch stellar models within the hot-flasher scenario. Our computations predict a new instability strip for hot-subdwarf stars centred in the $\log T_{\rm eff}-\log g$ diagram at slightly lower gravities than the canonical hot-subdwarf pulsators.  The locus of the instability domain is roughly $22000\,{\rm K} \lesssim T_{\rm eff}\lesssim 50000\,$K and $4.67 \lesssim \log g \lesssim 6.15$. The range of excited periods is $P \sim 200-2000\,$s corresponding to low/intermediate-order $g$ modes. Consequently, our computations show the excitation of long-period $g$ modes driven by the $\epsilon$ mechanism associated to pre-EHB stellar models, but not to
pre-BHB stellar models.

We found that a considerable number of excited modes are very likely to grow to observable amplitudes during the subflashes.  
Also, the rates of period change predicted are very high ($1-300\,$s/yr). These features shows that $\epsilon$-mechanism driven pulsations in hot-subdwarf stars are possible to be detected and identified.

The location of known He-rich hot subdwarf pulsators (Fig. \ref{fig:kiel}) is well reproduced by our theoretical predictions, although the observed periods in these stars are systematically longer than the predicted ones. In the cases of the longest observed periods in KIC 1718290 and UVO 0825+156 the pulsations cannot be  explained by $\epsilon$ mechanism due to the strong radiative damping of high-order $g$ modes in the stellar core. For LS IV-14$^\circ$116, our computations are able to explain the shortest observed period as well as its location in the $\log T_{\rm eff}-\log g$ diagram and its He-enriched composition. 
For this star it was proposed that a magnetic field could be involved in its variability \citep{2011MNRAS.412..363N,2011ApJ...734...59G}. But \citet{2015A&A...576A..65R} searched for a magnetic field in LS IV-14$^\circ$116 and found no evidence of it. Therefore, we conclude that the $\epsilon$ mechanism remains the best available explanation for the pulsations in this star. Although, as only the shortest period can be explained it is necessary to continue exploring alternative scenarios. 

\cite{2013EPJWC..4304004M} showed that $\epsilon$ mechanism can excite pulsations during the off-centred He-shell flashes that take place after a double He-white dwarf merger. The merger of two He-WD is also a very favourable scenario for the formation of He-sdO stars \citep{2012MNRAS.419..452Z}. Therefore, we intend to study in future works the properties of $\epsilon$-mechanism excited pulsations in detailed double He-WD post-merger models.

Finally, we conclude that our work could constitute a theoretical basis for future searches of pulsators in the Galactic field. In particular, based on simple numerical estimates we expect 1 to 10 stars in the current samples of hot-subdwarf stars to be pulsating by the $\epsilon$ mechanism.

\begin{acknowledgements}
The authors thank the anonymous referee for her/his suggestions that improved the original version of the article. This project has been partially supported by ANPCyT through grant PICT-2014-2708, MinCyT-DAAD bilateral cooperation through grant DA/16/07 and by a Return Fellowship from the Alexander von Humboldt Foundation. This research has made use of NASA's Astrophysics Data System Bibliographic Services.
\end{acknowledgements}

   \bibliographystyle{aa} 
   \bibliography{bibliografia} 

\end{document}